\documentclass[lettersize,journal]{IEEEtran}

\usepackage[switch]{lineno}
\IEEEoverridecommandlockouts

\usepackage{tipa}
\usepackage{amsmath,amsfonts}
\usepackage{algorithmic}
\usepackage{algorithm}
\usepackage{array}
\usepackage{adjustbox}
\usepackage[caption=false,font=normalsize,labelfont=sf,textfont=sf]{subfig}
\usepackage{textcomp}
\usepackage{stfloats}
\usepackage{url}
\usepackage{verbatim}
\usepackage{graphicx}
\usepackage{multirow, threeparttable, booktabs, makecell}
\usepackage{amssymb} 
\usepackage{pifont}  
\usepackage[table]{xcolor}
\usepackage{color}
\usepackage{enumitem}
\usepackage{placeins}
\usepackage{afterpage}
\usepackage{hyperref} 
\usepackage[numbers,sort&compress]{natbib}

\begin{document}
\bstctlcite{bstctl:etal, bstctl:nodash, bstctl:simpurl}

\title{PMF-CEC: Phoneme-augmented Multimodal Fusion for Context-aware ASR Error Correction with Error-specific Selective Decoding}

\author{Jiajun He and Tomoki Toda,~\IEEEmembership{Senior Member, IEEE}
\thanks{This work was supported in part by JST CREST Grant Number JPMJCR22D1, Japan, and a project, JPNP20006, commissioned by NEDO.}
\thanks{Jiajun He is with the Graduate School of Informatics, Nagoya University, Nagoya 464-8601, Japan (e-mail: \href{mailto:jiajun.he@g.sp.m.is.nagoya-u.ac.jp}{jiajun.he@g.sp.m.is.nagoya-u.ac.jp}).}
\thanks{Tomoki Toda is with the Information Technology Center, Nagoya University, Nagoya 464-8601, Japan (e-mail: \href{mailto:tomoki@icts.nagoya-u.ac.jp}{tomoki@icts.nagoya-u.ac.jp}).}
}

\markboth{Journal of \LaTeX\ Class Files,~Vol.~14, No.~8, August~2021}%
{Shell \MakeLowercase{\textit{et al.}}: A Sample Article Using IEEEtran.cls for IEEE Journals}


\maketitle

\begin{abstract}
End-to-end automatic speech recognition (ASR) models often struggle to accurately recognize rare words.
Previously, we introduced an ASR postprocessing method called error detection and context-aware error correction (ED-CEC), which leverages contextual information such as named entities and technical terms to improve the accuracy of ASR transcripts. Although ED-CEC achieves a notable success in correcting rare words, its accuracy remains low when dealing with rare words that have similar pronunciations but different spellings.
To address this issue, we proposed a phoneme-augmented multimodal fusion method for context-aware error correction (PMF-CEC) method on the basis of ED-CEC, which allowed for better differentiation between target rare words and homophones.
Additionally, we observed that the previous ASR error detection module suffers from overdetection. To mitigate this, we introduced a retention probability mechanism to filter out editing operations with confidence scores below a set threshold, preserving the original operation to improve error detection accuracy.
Experiments conducted on five datasets demonstrated that our proposed PMF-CEC maintains reasonable inference speed while further reducing the biased word error rate compared with ED-CEC, showing a stronger advantage in correcting homophones. Moreover, our method outperforms other contextual biasing methods, and remains valuable compared with LLM-based methods in terms of faster inference and better robustness under large biasing lists.
\end{abstract}

\vspace{-3mm}

\section{Introduction}
\IEEEPARstart{I}n recent years, end-to-end (E2E) automatic speech recognition (ASR) models have made significant strides across various domains, surpassing conventional hybrid ASR models \cite{radford2022robust}. However, even state-of-the-art (SOTA) E2E ASR systems encounter difficulties when transcribing domain-specific rare words or phrases, such as named entities, technical terms, and personal names from contact lists. These rare words, owing to their low frequency or absence in training data, are often incorrectly transcribed as more common words with similar pronunciations from the vocabulary \cite{huang24f_interspeech}.
ASR transcripts frequently serve as inputs for downstream spoken language processing (SLP) tasks, including video text summarization \cite{yang2024multi}, named entity recognition \cite{zhang2024context}, and emotion recognition \cite{he2024mf}. Incorrect transcription of rare words significantly impacts the accuracy of downstream models, thereby reducing the overall performance of the SLP system.

\begin{figure}[t]
  \centering
  \includegraphics[width=1\columnwidth]{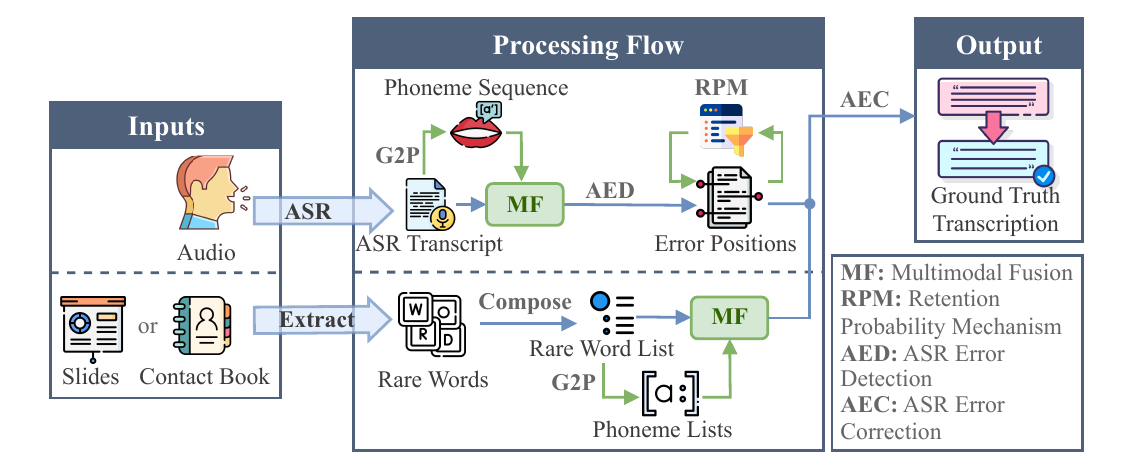}
  \vspace{-8mm}
  \caption{Illustration of the differences between our previous ED-CEC method and our proposed PMF-CEC method. 
   Shared components are marked with blue arrows, and new contributions in PMF-CEC are highlighted in green.
   }
  \vspace{-4mm}
  \label{fig:pipeline}
  \vspace{-3mm}
\end{figure}

To tackle these challenges, several context-aware approaches based on E2E ASR systems have been proposed. These methods leverage contextual knowledge to rectify ASR outputs, often by compiling rare word lists (or biasing lists) sourced from lecture slides and videos, meeting transcripts, and user contact lists \cite{sun2022tree, yang24f_interspeech}. Some approaches integrate additional language models (LMs) during decoding to adjust recognition results on the basis of contextual knowledge \cite{williams2018contextual, le2021deep, chen2022factorized}. However, these methods often require careful adjustment of interpolation parameters after ASR model updates. Others employ Trie structures to augment model constraints and contextual biasing during decoding, a technique called deep biasing (DB) \cite{sun2021tree, sun2022tree, sun2024graph}. However, these methods are resource-intensive and susceptible to errors when rebuilding or updating Trie structures upon changes in rare word lists. Alternatively, some methods modify ASR model architectures, such as directly integrating contextual information into the encoder, to enhance recognition accuracy \cite{pundak2018deep, huber2021instant, huang24f_interspeech}. 


Recent work has explored LLM-based contextual biasing by leveraging the strong language modeling and contextual reasoning capabilities of large language models (LLMs) \cite{bai2024seed, yang24f_interspeech}. These approaches typically consist of a speech encoder, a trainable adapter module and a decoder-only LLM that jointly processes contextual prompts and acoustic representations \cite{ma2024embarrassingly}. By injecting rare or domain-specific terms into the prompt as biasing lists, the model can generate more accurate transcriptions.
However, its performance is highly sensitive to prompt design and suffers from high inference latency and limited precision when handling large biasing lists.

To further enhance ASR system performance, researchers have proposed ASR error correction (AEC) methods \cite{zhang2023patcorrect, wang2022towards, DBLP:journals/corr/abs-2208-04641, li2024crossmodal}. These methods not only enable the postprocessing of ASR outputs but also provide several advantages. First, AEC methods do not require any modifications to existing ASR acoustic models, making them compatible with any ASR system. Additionally, AEC methods can be flexibly deployed at the system level, effectively improving the overall ASR accuracy without the need to retrain or adjust the underlying models. Notably, Wang et al. \cite{wang2022towards} proposed integrating contextual knowledge into error correction models via a context encoder, effectively enhancing ASR output quality. Nevertheless, this method introduces significant latency in inference speed.
Our prior work, ED-CEC \cite{he2023ed}, introduced an ASR error detection (AED) module to pinpoint error locations and perform context-aware error correction selectively. This approach notably accelerates inference speed and excels in handling rare words; yet, it may encounter challenges with rare words that are phonetically similar but spelled differently.

Recent studies underscore the pivotal role of phonetic information in AEC \cite{zhang2022asr, he2023enhancing,  zhang2023patcorrect, dong2024pronunciation}. For instance, Zhang et al. \cite{zhang2023patcorrect} achieved substantial improvements in AEC accuracy by fusing phonetic information with textual representations through multimodal fusion. This underscores the advantage of combining phonetic information over text-only postprocessing models, prompting us to integrate phoneme information into context-aware AEC, particularly when ASR transcriptions lack sufficient information for biasing correction.

Beyond conventional methods, LLM-based generative approaches have recently gained traction for AEC \cite{chen2023hyporadise, yang2023generative, yang2024large, hu2024large, li2025revise, ma2025asr}. These models effectively use LLMs to directly revise ASR outputs, either by reranking N-best hypotheses or generating corrected transcriptions in a free-form manner. Despite their potential, such approaches typically rely on fully autoregressive decoding with beam search, leading to a high computational cost.
 They also suffer from overcorrection, limited control over edit granularity, and require considerable training data and computational resources to align audio with textual inputs. These limitations hinder their applicability in real-time or resource-constrained scenarios, underscoring the need for lightweight, efficient, and controllable alternatives.

Therefore, building on our previous ED-CEC work, in this paper, we propose a novel phoneme-augmented multimodal fusion approach to context-aware error correction (PMF-CEC), enhancing traditional input text encoding with phoneme augmentation to correct ASR errors in rare words that are phonetically similar but spelled differently. The specific differences between ED-CEC and PMF-CEC are shown in Fig. \ref{fig:pipeline}. Specifically, PMF-CEC enhances ED-CEC by incorporating phoneme-augmented multimodal fusion, enabling better differentiation between rare words and homophones. To address the overdetection issue in the AED module, we introduce a retention probability mechanism (RPM) that filters out low-confidence edits during the inference stage, improving detection accuracy. Our contributions are summarized as follows:

\noindent
\begin{itemize}[leftmargin=*]
\setlength{\topsep}{0pt}
\setlength{\itemsep}{0pt}
\setlength{\parsep}{0pt}
\setlength{\parskip}{0pt}
\item \noindent We proposed a partially autoregressive PMF-CEC model based on the Transformer architecture, enhancing traditional input text encoding through multimodal fusion to incorporate phonetic information for improved AEC.
\item We introduced the RPM during inference to improve error detection accuracy. In the AED module, confidence estimation was conducted for each editing operation (Keep, Delete, and Change), allowing the model to filter out low-confidence edits and make more precise corrections.
\item Extensive experiments on five public datasets demonstrated that PMF-CEC outperforms ED-CEC using text-only modality, achieving 4.25\% to 14.46\% in word error rate reduction (WERR) and 8.39\% to 12.56\% in biased word error rate (B-WER). This validates the effectiveness of PMF-CEC in correcting phonetically similar but differently spelled rare words. Moreover, our proposed method maintains inference speed within the same range of 27.21 to 38.12 milliseconds (ms) while being approximately 2.4 to 5.3 times faster than the SOTA autoregressive model.
\item We further compared PMF-CEC with representative LLM-based ASR models and show that, while achieving comparable or even superior correction performance, our method offered significantly faster inference and greater robustness under large biasing lists. This makes PMF-CEC a more practical and scalable solution for real-time ASR applications.
\end{itemize}


\section{Related Work}

Recently, various contextual biasing methods have been developed for E2E ASR, primarily including graph fusion methods, attention-based deep context approaches, and AEC. 
In addition, the rapid advancements in LLMs have opened new possibilities for improving contextual biasing in ASR systems.
In this section, we provide a brief overview of traditional contextual biasing methods as well as recent efforts to incorporate LLMs into ASR.

\vspace{-3mm}
\subsection{Graph Fusion Methods}
Graph fusion methods primarily involve approaches based on finite state transducers (FSTs) and Trie-based methods, as well as their combination. On one hand, contextual biasing based on FST models has demonstrated effectiveness in both traditional \cite{aleksic2015bringing, mcgraw2016personalized} and E2E \cite{zhao2019shallow, williams2018contextual} ASR systems. These methods utilize LM interpolation or shallow fusion (SF), which are applicable even with minimal training data \cite{tang2024improving, munkhdalai2023nam+}. FST-based contextual biasing allows precise control over token transitions through weighted schemes, enabling ASR systems to incorporate domain-specific knowledge about the distribution of rare words. However, traditional FST biasing significantly complicates the inference process, especially in E2E ASR systems. Moreover, since traditional FST models cannot be co-optimized with ASR models via gradient descent, careful readjustment of interpolation parameters is typically necessary following ASR model updates, posing challenges when managing multiple FST models. Additionally, these methods are often limited to specific contexts, such as ``call [contact name]" and ``play my [playlist] on Spotify", which restricts their capability to handle diverse grammatical structures in natural language \cite{fu2023robust, sun2021tree}.

On the other hand, Trie-based contextual biasing methods use Trie structures to enhance model constraints and contextual biases during the decoding process, a technique called DB. This approach leverages the efficient organization and retrieval capabilities of Tries to help ASR models more accurately predict and select vocabulary. Le et al. \cite{le2021contextualized} explored combining Trie-based methods with SF using WFST, extending this integration to RNN LMs to improve the handling of biasing words. They increased efficiency by extracting biasing vectors from Trie constraints representing biasing lists. Additionally, Sun et al. \cite{sun2021tree, sun2022tree} proposed the tree-constrained pointer generator component (TCPGen), which uses a structured Trie representation of biasing words to create a neural shortcut between the biasing lists and the final model output distribution. This method effectively addresses the challenge of managing large biasing lists \cite{sun2024graph}.

\vspace{-4mm}
\subsection{Attention-based Deep Context Approaches}
Recently, owing to the improved recognition of rare words and the ease of integration with E2E neural inference engines, context biasing methods based entirely on neural attention mechanisms have gained increasing popularity. Neural context biasing methods in LAS have been discussed in \cite{pundak2018deep}, where biasing phrases and contextual entities are encoded via BiLSTM encoders and biased using a position-aware attention mechanism \cite{chen2019joint, bruguier2019phoebe}. 
For RNN-T models, in \cite{chang2021context, sathyendra2022contextual}, fully neural attention-based context biasing methods were introduced. In \cite{chang2021context, sathyendra2022contextual}, context biasing methods for Transformer Transducers (T-T) and Conformer Transducers (C-T) in auditory encoders and text-based prediction networks were introduced. In \cite{sathyendra2022contextual}, the use of context adapters to adapt pretrained RNN Transducer (RNN-T) and C-T models was discussed, which have the advantages of being faster and more data-efficient. Sudo et al. \cite{sudo2024contextualized} combined biasing phrase index loss and special token training to enhance context performance during inference using the biasing phrase boosted beam search algorithm. Yu et al. \cite{yu2024lcb} proposed LCB-net, employing dual encoders to model audio and long-context biasing, and enhancing model generalization and robustness through dynamic context phrase simulation. Although attention-based methods eliminate the dependence on syntactic prefixes seen in SF methods, they require more memory during training and inference, alter the structure of the original ASR model, which potentially leads to decreased performance of the original ASR model, and are less effective in handling large biasing lists \cite{sun2022tree}.

\vspace{-3mm}

\begin{table*}[ht]
\centering
\caption{Comparison of several contextual AEC methods with the proposed method}
\vspace{-3mm}
\label{table:realted_word}
\resizebox{\linewidth}{!}{
\begin{tabular}{lcccc}
\toprule
\textbf{Method}                     & \textbf{Handles Homophones} & \textbf{Fast Inference} & \textbf{Avoids Overcorrection} & \textbf{Scalable to Large Biasing Lists} \\
\midrule
\midrule
CSC \cite{wang2021light}               & No                          & No                     & No                             & Yes                                 \\
Jiang et al. \cite{jiang2024contextual}         & No                          & No                      & No                              & Yes                                \\
FCSC \cite{wang2022towards}             & No                          & Yes                     & Yes                             & Yes                                \\
PGCC \cite{dong2024pronunciation}       & Yes                         & No                      & Yes                             & No                                 \\
\midrule
\midrule
ED-CEC \cite{he2023ed}    & No                          & Yes                     & Yes                             & Yes                                 \\
\rowcolor[HTML]{EFEFEF}
\textbf{PMF-CEC}        & \textbf{Yes}                         & \textbf{Yes}                     & \textbf{Yes}                             & \textbf{Yes}                                \\
\bottomrule
\end{tabular}
}
\vspace{-3mm}
\end{table*}

\subsection{ASR Error Correction (AEC)}
AEC is proven effective in refining errors generated by ASR models, thereby significantly reducing the word error rate (WER) at the ASR postprocessing stage. Wang et al. \cite{wang2021light} introduced a lightweight contextual spelling correction model (CSC) to rectify context-related recognition errors in transcription-based ASR systems, utilizing a shared context encoder and filtering algorithms for large biasing lists. For document-level AEC, Jiang et al. \cite{jiang2024contextual} proposed a kNN-based context-aware model with enhanced performance through retrieval from context data stores. Methods for AEC based on autoregressive sequence-to-sequence architectures may suffer from overcorrection issues, introducing new errors or alterations to correct portions and causing significant inference latency. To address these challenges, Wang et al. \cite{wang2022towards} integrated context information into a non-autoregressive spelling correction model with a shared context encoder, while introducing filtering algorithms and performance balancing mechanisms to control the biasing extent for large biasing lists.
Dong et al. \cite{dong2024pronunciation} proposed the pronunciation guided copy and correction (PGCC) model for AEC, leveraging the encoder-decoder structure of BART pretraining to optimize decisions on whether to copy source input tokens or generate modified ones, effectively identifying and correcting homophone errors.

The proposed PMF-CEC method stands out among existing AEC approaches in terms of integrating phoneme-augmented multimodal fusion into context-aware error correction, which addresses the limitations of text-only models. As shown in Table \ref{table:realted_word}, compared with CSC \cite{wang2021light} and FCSC \cite{wang2022towards}, PMF-CEC not only preserves the efficiency and low latency of non-autoregressive models but also excels at correcting phonetically similar but differently spelled rare words, a common issue in ASR. Although PGCC \cite{dong2024pronunciation} focused on homophone errors, PMF-CEC provides a more robust solution through a combination of phonetic and textual information. The RPM further refines the correction process, ensuring both accuracy and inference speed. 
\vspace{-3mm}

\subsection{LLM-based ASR and AEC methods}

In recent years, with the development of LLMs, LLM-based approaches for ASR contextual biasing and AEC have attracted increasing attention. These methods leverage the strong contextual modeling and multimodal reasoning capabilities of LLMs by integrating frozen speech encoders with prompt-based frameworks. For example, Seed-ASR \cite{bai2024seed} and MaLa-ASR \cite{yang24f_interspeech} incorporate domain knowledge or presentation keywords as prompts, significantly improving recognition of rare and domain-specific words while reducing WER.

Meanwhile, generative LLMs have been widely applied to AEC tasks. HyPoradise \cite{chen2023hyporadise} introduces an LLM-based benchmark for N-best list correction, demonstrating the ability to recover missing hypotheses beyond traditional reranking. Yang et al. further propose task-activating prompting techniques \cite{yang2023generative} and extend the scope to speaker attribution and emotion recognition in the GenSEC challenge \cite{yang2024large}. In addition, Hu et al. \cite{hu2024large} explore robustness under noisy conditions by introducing noise-aware embeddings in the language space.

Despite promising results, LLM-based methods still face several practical challenges. They are computationally intensive, suffer from high inference latency, and are sensitive to prompt design. Their performance also degrades significantly when handling large-scale biasing vocabularies. Moreover, incorporating acoustic features into LLM training leads to increased training costs, and current lightweight adaptations still rely heavily on large-scale data with limited generalization. These limitations suggest that relying solely on LLMs is possibly insufficient for robust and scalable contextual biasing and correction in real-world ASR applications.

\begin{figure*}[!t]
  \centering
  \includegraphics[scale=0.13]{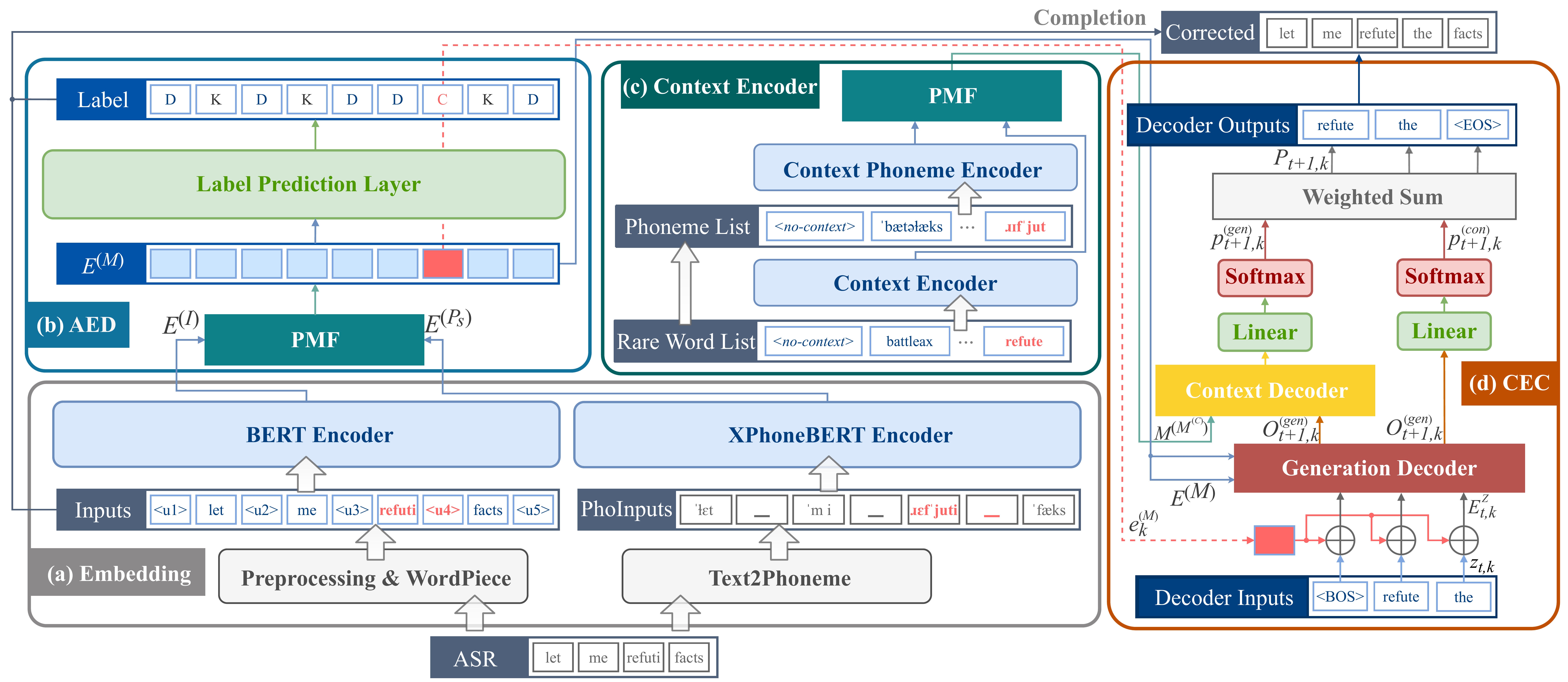}
  \vspace{-3mm}
  \caption{Overall architecture of the proposed PMF-CEC model.}
  \label{fig:model}
  \vspace{-5mm}
\end{figure*}

\section{Proposed Methodology}

\subsection{Problem Formulation}
\label{section2.0}
The contextual error correction problem can be formalized as the mapping function $f(S,C) = T$, where the source $S = (s_1, s_2, \cdots , s_m) \in \mathbb{R}^{m} $ represents the original ASR transcript, the context $C = (C_1, C_2, \cdots , C_l) \in \mathbb{R}^{l}$ denotes the rare word list containing $l$ contextual items, and the target $T = (t_1, t_2, \cdots , t_n) \in \mathbb{R}^{n}$ is the ground truth transcript. All the tokens are applied to a predefined WordPiece vocabulary \cite{wu2016google}. Additionally, to incorporate multimodal information, we included the phoneme sequences: $P^{s} = (p^s_1, p^s_2,\cdots , p^s_q) \in \mathbb{R}^{q}$ for the source sentence $S$, and $P^{C} = (P^{(C)}_1, P^{(C)}_2,\cdots , P^{(C)}_l) \in \mathbb{R}^{l}$ for the context $C$, where $q$ denotes the length of phonemes in the source sentence.

\vspace{-4mm}
\subsection{Preprocessing}
\label{section2.1}

Similarly to the preprocessing described in \cite{DBLP:journals/corr/abs-2208-04641, he2023ed}, we first inserted dummy tokens (e.g., $<$u1$>$, $<$u2$>$, etc.) between every two consecutive words in $S$, as shown in Fig. \ref{fig:model}(a). This insertion allows us to directly add new tokens in the gaps between existing tokens, thereby reducing the ambiguity of possible editing operations. Consequently, the original tokens can only be retained or deleted, whereas dummy tokens can only be deleted or replaced with other tokens.
In addition, the phoneme sequences for each word were generated using the \texttt{Text2PhonemeSequence} tool\footnote{\url{https://github.com/thelinhbkhn2014/Text2PhonemeSequence}}.

To construct the supervised labels for the error detection module, we defined three editing operations: \textit{KEEP} (\textbf{K}), \textit{DELETE} (\textbf{D}), and \textit{CHANGE} (\textbf{C}). Intuitively, \textbf{K} means that the token should be retained, \textbf{D} indicates that the token should be removed, and \textbf{C} signifies that the token should be replaced with another token.

By determining the longest common subsequence (LCS) between $S$ and $T$, we aligned $S$ and $T$. With the inserted dummy tokens, we generated an aligned representation of $S$, denoted as $I = (i_1, i_2, \cdots, i_{2m+1}) \in \mathbb{R}^{2m+1}$, where $m+1$ represents the number of inserted dummy tokens. Finally, the aligned tokens were labeled \textbf{K}, the positions requiring correction were labeled \textbf{C}, and the remaining tokens were labeled \textbf{D}. An example of this process is illustrated in Fig. \ref{fig:model}(b). It is clear that compared with the fully autoregressive AEC, our proposed method only requires adding tokens at positions needing correction, whereas other positions can remain unchanged or be deleted directly, significantly improving inference speed. Additionally, Table \ref{tab:avg_len} compares the average length between the added tokens and the ground truth transcriptions, demonstrating that the introduction of editing operations saves at least four-fifths of the decoding time compared with fully autoregressive AEC models.

\begin{table}[t]
    \centering
    \caption{Comparison of the average length between the added tokens and the ground truth transcripts on Librispeech test sets.}
    \label{tab:avg_len}
    \begin{tabular}{l|cc}
        \toprule
        & \textbf{Added Tokens} & \textbf{Ground Truth Transcripts} \\
        \midrule
        \midrule
        \textbf{Avg. length} & 3.00 & 18.94 \\
        \bottomrule
    \end{tabular}
    \vspace{-3mm}
    
\end{table}

\vspace{-3mm}
\subsection{Embedding Module}
\label{section2.2}
Our embedding module consists of a text encoder and a phoneme encoder used to obtain the contextual token representations and the contextual phoneme representations, respectively. 
Next, we provide a detailed explanation of each encoder.

\textbf{Contextual Token Representations.} 
We employed the pretrained LM BERT \cite{devlin2018bert}, a multilayer bidirectional Transformer encoder \cite{vaswani2017attention}, as our text encoder to obtain the contextual token representations $E^{(I)} = (e^{(I)}_1, e^{(I)}_2, \cdots , e^{(I)}_{2m+1}) \in \mathbb{R}^{(2m+1) \times d_h}$ for the text inputs $I$, where $d_h$ denotes the output dimension:
\begin{equation}
    E^{(I)} = {\rm BERT}({\rm TE}(I)+{\rm PE}(I)),
\label{eq1}
\end{equation}
where TE and PE denote the token embedding and position embedding, respectively.

\textbf{Contextual Phonetic Representations.} To acquire comprehensive contextual phonetic representations $E^{(P_s)} = (e^{(P_s)}_1, e^{(P_s)}_2, \cdots , e^{(P_s)}_{q}) \in \mathbb{R}^{q \times d_h}$ for the phoneme inputs $P_s$, we leveraged a pretrained self-supervised learning (SSL) model, XPhoneBERT \cite{thenguyen23_interspeech} as our phoneme encoder. 
XPhoneBERT has the same model architecture as BERT, with 12 Transformer blocks, a hidden size of 768, and 12 self-attention heads.

\vspace{-5mm}
\subsection{Phoneme-augmented Multimodal Fusion (PMF) Module}

\begin{figure}[ht]
\vspace{-6mm}
  \centering
  \includegraphics[width=0.4\columnwidth]{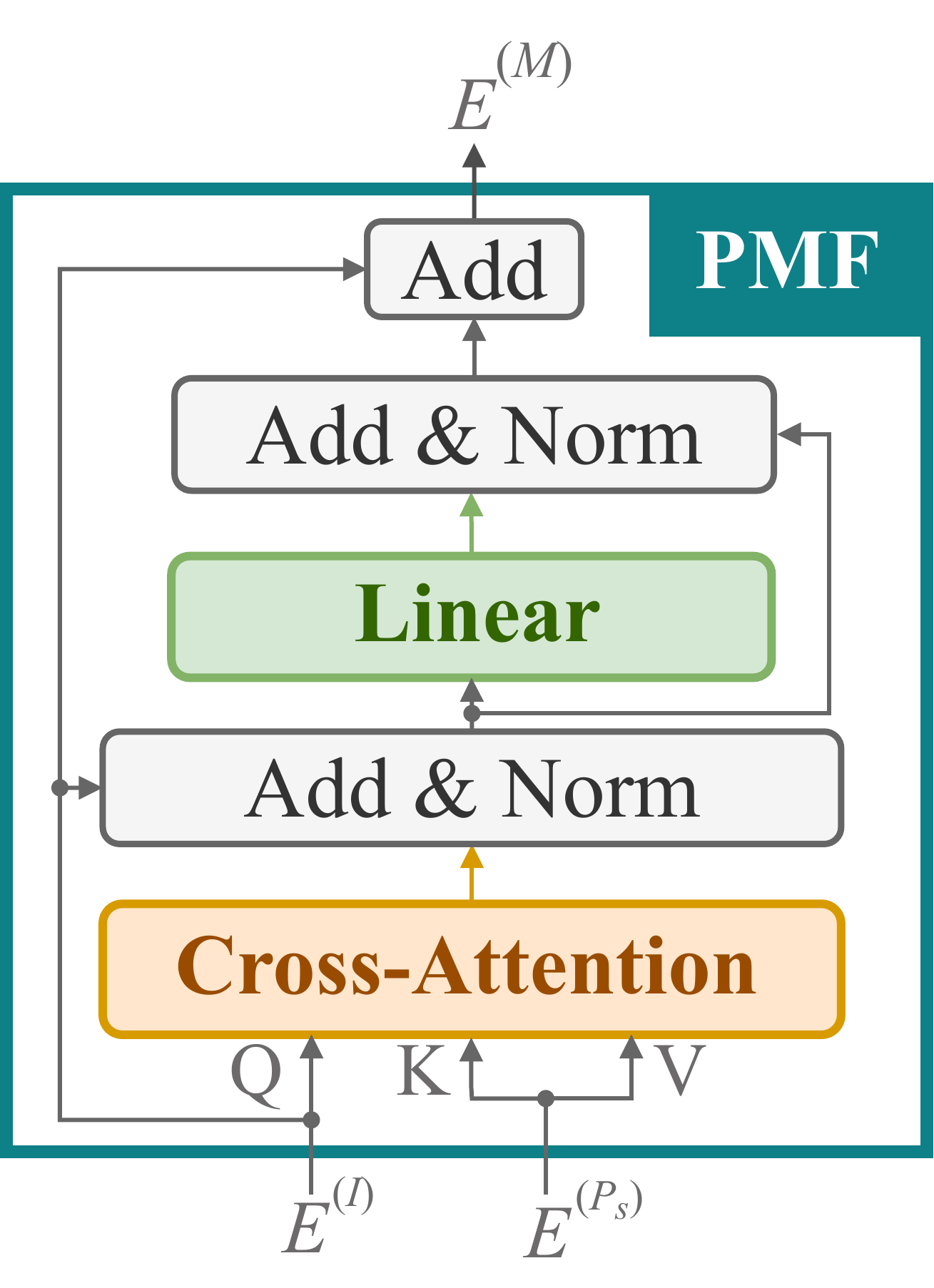}
  \vspace{-4mm}
  \caption{Illustration of the PMF module.}
  \vspace{-6mm}
  \label{fig:PMI}
\end{figure}

To address the varying embedding lengths across different modalities, we incorporated a cross-attention layer in the text encoder's output, facilitating alignment with the output from the phoneme encoder.
The cross-attention outputs were computed using $E^{(I)}$ as the query and $E^{(P_s)}$ as both the key and the value:

\vspace{-3mm}
\begin{equation}
\label{eq2}
(E^{(M)})^{'} = \text{Softmax}\left( \frac{(E^{(I)})(E^{(P_s)})^\top}{\sqrt{d_h}} \right)(E^{(P_s)}),
\end{equation}
where $(E^{(M)})^{'} \in \mathbb{R}^{(2m+1) \times d_h}$ is the output after additional linear transformation, residual connection, and normalization layers. Then, we combined the weighted text-aware phoneme representation and the original token representation using the addition operation to obtain the final multimodal representation $E^{(M)} = (e^{(M)}_1, e^{(M)}_2, \cdots , e^{(M)}_{2m+1}) \in \mathbb{R}^{(2m+1) \times d_h}$, as illustrated in Fig. \ref{fig:PMI}:
\begin{equation}
\label{eq3}
E^{(M)} = (E^{(M)})^{'} + E^{(I)}.
\end{equation}

\subsection{ASR Error Detection (AED) Module}
The AED module takes the multimodal representation $E^{(M)}$ as input to predict editing labels for each token in the original ASR transcript, as shown in Fig. \ref{fig:model}(b).
%
The label prediction layer consists of a simple fully connected network designed to output classification probability scores for three categories: \textbf{K}, \textbf{D}, and \textbf{C}. Owing to its simplicity, the error detection module has a negligible impact on the overall model size but significantly enhances the system's inference speed.

\begin{equation}
\label{eq4}
  P(y_o|e_o) = {\rm Softmax}({\rm FC}(e_o)) \in \mathbb{R}^{3},
\end{equation}
where $e_o \in E^{(M)}$ and $y_o$, $o \in \{1, \cdots, 2m+1\}$ are the multimodal representation and predicted labeling operations of the $o^{th}$ token, respectively. $\rm{FC}$ is a fully connected layer.

\vspace{-3mm}
\subsection{Context-aware Error Correction (CEC) Module}

In contrast to conventional autoregressive decoders that initiate error correction from the ground up, our CEC module concurrently processes all \textbf{C} tokens identified by the AED module. This module either generates new tokens via the transformer decoder or selects pertinent tokens from the rare word list to rectify ASR errors, thereby optimizing the correction process, as denoted in Figs. \ref{fig:model}(c) and \ref{fig:model}(d).


\textbf{Decoder Inputs.} For the $k^{th}$ \textbf{C} position, the generated decoding sequence of length $T$ by the transformer decoder can be represented as $Z_k = (z_{1,k}, z_{2,k}, \cdots , z_{T,k}) \in \mathbb{R}^{T}$,  where $z_{1,k}$ is initialized by a special start token $<$\textit{BOS}$>$. We computed the decoder inputs at step $t$ as follows:
\begin{equation}
  E_{t,k}^{(Z)} = {\rm FC}(({\rm TE}(z_{t,k})+{\rm PE}(z_{t,k})) \oplus e^{(M)}_k \in \mathbb{R}^{d_h},
\end{equation}
where $e^{(M)}_k$ is the multimodal representation of the $k^{th}$ change position. ``$\oplus$" denotes a concatenate function. $\rm{FC}$ is the fully connected layer that maps the decoder inputs back to the same dimension as the embedding of $z_{t,k}$.

\textbf{Generation Decoder.} 
To generate new tokens, we used the output of the decoder input $\hat{E}_{t,k}^{(Z)}$ as the query and the multimodal representation $E^{(M)}$ as the key and value to the Transformer decoder to obtain the output representation of the decoder layer:
%
%
%
%
\begin{equation}
\label{eq6}
O_{t+1,k}^{(gen)} = \text{Softmax}\left( \frac{(\hat{E}_{t,k}^{(Z)})(E^{(M)})^\top}{\sqrt{d_h}} \right)(E^{(M)}),
\end{equation}
where $O_{t+1,k}^{(gen)} \in \mathbb{R}^{d_h}$ is the decoder layer output after additional linear transformation, residual connection, and normalization layers. 
Finally, the generation output was calculated as
%

\begin{equation}
p_{t+1,k}^{(gen)} = {\rm Softmax}({\rm FC}(O_{t+1,k}^{(gen)})) \in \mathbb{R}^{d_{vb}},
\end{equation}
where $d_{vb}$ is the vocabulary size of the BERT.  Therefore, the next generated token is $z_{t+1,k} = {\rm argmax}(p_{t+1,k}^{gen})$. 

%

Additionally, we introduced a context mechanism that dynamically selects between generating new tokens through the generation decoder or choosing relevant tokens from a preprepared rare word list. This context mechanism consists of a context encoder and a context decoder, with the context decoder comprising context attention and context-item attention. The detailed description is as follows:

\textbf{Context Encoder.} 
We stored $l$ contextual items, consisting of rare words or phrases, in the rare word list, with the construction process detailed in Section \ref{sec:rw_con}. 
The  $j^{th}$ contextual item and the corresponding phoneme sequence are represented as $C_j = (c_j^1, \cdots, c_j^u) \in \mathbb{R}^{u}$ and $P^{(C)}_j = ((p^{(c)}_j)^1, \cdots, (p^{(c)}_j)^v) \in \mathbb{R}^{v}$, $j \in \{1, 2, \cdots, l\}$, where $u$ and $v$ denote the numbers of tokens and phonemes in the $j^{th}$ contextual item, respectively.
To optimize model size and improve inference speed, we implemented parameter sharing between the BERT encoder and the contextual encoder. Thus, we used the same BERT encoder to obtain the token representations of each contextual item $E^{(C_j)} = (e_{1}^{(c_j)}, e_{2}^{(c_j)}, \cdots, e_{u}^{(c_j)}) \in \mathbb{R}^{u \times d_h}$:
%
\begin{equation}
E^{(C_j)}={\rm BERT}({\rm TE}((C_j))+{\rm PE}((C_j))),
\end{equation}
where $C_j$ is the $j^{th}$ contextual item.
Similarly, 
we also obtained the corresponding phoneme sequences $E^{(P^{(C)}_j)} = (e_{1}^{(p^{(c)}_j)}, e_{2}^{(p^{(c)}_j)}, \cdots, e_{v}^{(p^{(c)}_j)}) \in \mathbb{R}^{v \times d_h}$ for all contextual items, as well as the multimodal representations $ E^{(M^{(C)}_j)} = (e_{1}^{(M^{(C)}_j)}, e_{2}^{(M^{(C)}_j)}, \cdots, e_{u}^{(M^{(C)}_j)}) \in \mathbb{R}^{u \times d_h}$. Therefore, for all contextual items, we defined them as $E^{(C)} = (E^{(C_1)}, E^{(C_2)}, \cdots, E^{(C_l)}) \in \mathbb{R}^{l \times u \times d_h}$, $E^{(P^{(C)})}= (E^{(P^{(C)}_1)}, E^{(P^{(C)}_2)}, \cdots, E^{(P^{(C)}_l)})  \in \mathbb{R}^{l \times v \times d_h}$, and $E^{(M^{(C)})} = (E^{(M^{(C)}_1)}, E^{(M^{(C)}_2)}, \cdots, E^{(M^{(C)}_l)}) \in \mathbb{R}^{l \times u \times d_h}$, respectively.

\begin{figure}[htb]
\vspace{-5mm}
\begin{minipage}[b]{1\linewidth}
  \centering
  \centerline{\includegraphics[width=6cm]{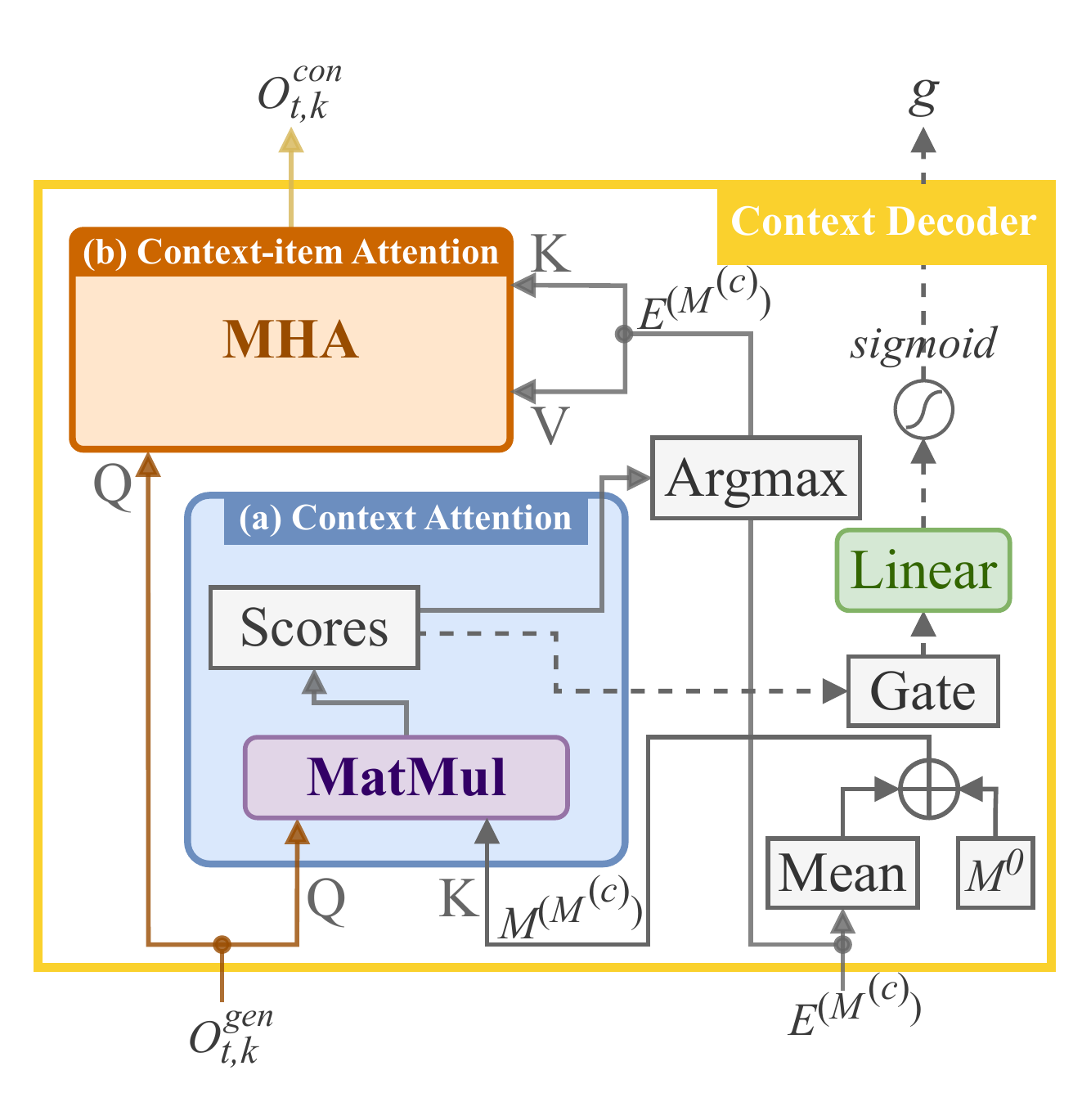}}
\end{minipage}
\vspace{-10mm}
\caption{Illustration of the context decoder.}
\label{fig:res}
%
\end{figure}

\textbf{Context Decoder.} 
The illustration of the context decoder is shown in Fig. \ref{fig:res}.
We first computed the average of the encoded contextual items and then introduced a learnable dummy token $<$\textit{no-context}$>$ at the beginning of these contextual items to determine whether relevant information is stored in the rare word list:
\begin{equation}
\Bar{E}^{(M^{(C)})} = \text{mean}(E^{(M^{(C)})})  \in \mathbb{R}^{l \times d_h},
\end{equation}
\begin{equation}
M^{(M^{(C)})} = M^0 \oplus \Bar{E}^{(M^{(C)})} \in \mathbb{R}^{(l+1) \times d_h},
\end{equation}
where $M^0 \in \mathbb{R}^{d_h}$ is the learned hidden representation of the dummy token $<$\textit{no-context}$>$
and $M^{(M^{(C)})}$ can be interpreted as the summarized tokens for each contextual item.

During step $t$, the contextual decoder first utilizes a context attention layer to determine the availability and specific positions of relevant contextual items from the rare word list. In this process, similarity scores are calculated by treating the output  $O_{t,k}^{(gen)}$ of the generation decoder as the query and summary contextual tokens $M^{(M_c)}$ as the key:
%
%
%


\begin{equation}
\text{scores}_t = O_{t,k}^{(\text{gen})} M^{(M^{(C)})^\top} \in \mathbb{R}^{(l + 1)},
\end{equation}

\begin{equation}
{\mbox gate}_t = \text{scores}_t^{(1)} \in \mathbb{R}^{1},
\end{equation}
where ${\mbox gate}_t$ are the similarity scores corresponding to the $<$\textit{no-context}$>$ token $M^0$.
We defined the index of the highest similarity score for the query $Q_t$ at step $t$ as $m = {\rm argmax}({\rm scores}_t)\in \mathbb{R}^{1}$. If $m$ is nonzero, indicating the presence of relevant contextual knowledge in the rare word list, we computed the contextual output using the context-item attention layer. This layer extracts the relevant information from a specific contextual item using a multihead attention (MHA) mechanism. In the MHA layer, the query input $Q_t$ is the output of the decoder layer $O_{t,k}^{(gen)} \in \mathbb{R}^{d_h}$, whereas the key input $K_t$ and the value input $V_t$ are both the multimodal representation $E^{(M^{(C)}_m)} \in \mathbb{R}^{u \times d_h}$ of the $m^{th}$ contextual item:
%
\begin{equation}
Q_t = O_{t,k}^{(gen)}, K_t = E^{(M^{(C)}_m)}, V_t = E^{(M^{(C)}_m)},
\end{equation}
%
%

%
\begin{equation}
O_{t+1,k}^{(con)} = {\rm Softmax}( \frac{Q_tK_t^{T}}{\sqrt{d_h}}  )V_t,
\end{equation}
\begin{equation}
p_{t,k}^{(con)} = {\rm Softmax}({\rm FC}(O_{t,k}^{(con)})) \in \mathbb{R}^{d_{vb}},
\end{equation}
where $O_{t,k}^{(con)} \in \mathbb{R}^{d_h}$ is the output of the context-item attention layer.
Note that no attention is calculated when the max score corresponds to the dummy token, namely, $m=0$. 
Then, the predicted word was acquired by a weighted sum between the generation output $p_{t,k}^{(gen)}$ and the contextual output $p_{t,k}^{(con)}$:
%
\begin{equation}
g =  \sigma ({\rm FC}({\mbox gate}_t)),
\end{equation}
%
%
\begin{equation}
P_{t,k} = g \cdot p_{t,k}^{(gen)} + (1 - g) \cdot p_{t,k}^{(con)},
\end{equation}
where $\sigma$ is the sigmoid function and $g$ is the gate to make a trade-off between the chosen token from the rare word list and the generated token by the generation decoder.

\vspace{-3mm}
\subsection{Joint Training}
The learning process was optimized through two objectives that correspond to error detection and context-aware error correction.
%
\begin{align}
{\rm Loss}_d & = - \sum_{o}{\rm log}(P(y_o|i_o)), \\
{\rm Loss}_e & = - \left( \sum_{k}\sum_{t}{\rm log}(P_{t,k}) + \right. \nonumber \\
              & \quad \left. \sum_{k}\sum_{t}{\rm log}(P({\rm label}_{t,k}|{\rm scores}_{t,k})) \right),
\end{align}
where the loss function ${\rm Loss}_d$ is the cross entropy loss for the detection network and the loss function ${\rm Loss}_e$ consists of two parts of the cross entropy loss for the context-aware correction network. Furthermore, ${\rm scores}_{t,k}$ are the score outputs of the context attention layer and ${\rm label}_{t,k}$ is the contextual label that contains the index of the corresponding contextual item. The two loss functions are linearly combined as the overall objective in the learning phase:
\vspace{-2mm}
%
\begin{equation}
{\rm Loss} = \gamma \cdot {\rm Loss}_d + {\rm Loss}_e,
\end{equation}
%
where $\gamma$ is the hyperparameter for adjusting the weight between ${\rm Loss}_d$ and ${\rm Loss}_e$.

\subsection{Inference}
During the completion process, we converted the predicted operation labels and the generated words into a complete utterance. Specifically, as depicted in Fig. \ref{fig:model}, we preserved the tokens labeled \textbf{K} and removed those labeled \textbf{D} from the inputs. We then replaced the tokens labeled \textbf{C} with the corresponding generated words.

The accuracy of error correction in our model is affected by the AED module. When the model overdetects, originally correct positions may be mistakenly identified as errors, leading to an increase in WER instead of a decrease. To address this issue, we introduced a RPM during the inference phase. We scored the confidence of each editing operation output by the AED. If the confidence is below the set threshold, the original editing operation at that position is retained, thereby reducing the risk of overdetection.
For example, in Fig. \ref{fig:model}, after inserting dummy tokens, the ASR-transcribed sentence is ``$<$u1$>$ let $<$u2$>$ me $<$u3$>$ refuti $<$u4$>$ facts $<$u5$>$”. The original editing operations should be ``D K D K D K D K D” without any modifications. After processing with the AED module, the editing operations are ``D K D K D D C K D”, with assumed confidence scores for each operation being ``0.8 0.6 0.7 0.9 0.8 0.4 0.3 0.6 0.7” and a set threshold of 0.5. Positions with confidence scores below 0.5 retain the original editing operations, resulting in the final editing operations being ``D K D K D K D K D”.

\section{Experimental Setup}
\label{sec:experiment_setup}
\subsection{Implementation Details}

Our method was implemented using Python 3.7 and Pytorch 1.11.0. The model was trained and evaluated on a system equipped with an Intel(R) Xeon(R) Gold 6248 CPU @ 2.50 GHz, 32 GB of RAM, and one NVIDIA Tesla V100 GPU. The detailed parameter settings are presented in Table \ref{tab:parameter_settings}.

\renewcommand{\arraystretch}{1.2} 

\begin{table}[ht]
\centering
\caption{Model configurations and hyperparameters in our experiments.}
\vspace{-3mm}
\label{tab:parameter_settings} 
\begin{tabular}{lc}
\toprule
\textbf{Configuration} & \textbf{Value}  \\
\midrule
\midrule
Epochs & 20 \\
Optimizer & Adam \\
Leraning rate & 0.00005 \\
Dropout & 0.1 \\
$d_h$ & 768 \\
Batch size & 32 \\
$\gamma$ & 3 \\

\midrule
\midrule
\rowcolor[HTML]{EFEFEF} 
\multicolumn{2}{c}{\textbf{Pretrained model}} \\
Word-based  & bert-base-uncased \\
$d_{vb}$ & 30,522 \\
\midrule
Phoneme-based  & xphonebert-base \\
$d_{vp}$ & 1,960 \\
\bottomrule
\end{tabular}
\end{table}

\renewcommand{\arraystretch}{1.5} 

Both the text encoder and the context encoder were initialized using the shared \texttt{bert-base-uncased}\footnote{\url{https://huggingface.co/google-bert/bert-base-uncased}} model, producing token representations with a dimensionality of 768. The vocabulary size for word tokenization, $d_{vb}$, was set to 30,522. Similarly, the phoneme encoder and the context phoneme encoder were initialized with the shared \texttt{xphonebert-base}\footnote{\url{https://huggingface.co/vinai/xphonebert-base}} model, generating phonetic representations with a dimensionality of 768. The vocabulary size for phoneme tokenization, $d_{vp}$, was set to 1,960. We configured the hidden size $d_h$ to 768, with 12 attention layers and 12 attention heads. The transformer decoder utilized a single-layer transformer with a hidden size of 768. We employed the Adam optimizer \cite{kingma2014adam} with a batch size of 32 and set $\gamma$ to 3. The initial learning rate was 0.00005. All hyperparameters were finetuned using standard validation data. Following empirical validation, we set the threshold of the RPM to 0.5 during inference.

\begin{figure*}[!t]
  \centering
  \includegraphics[width=2\columnwidth]{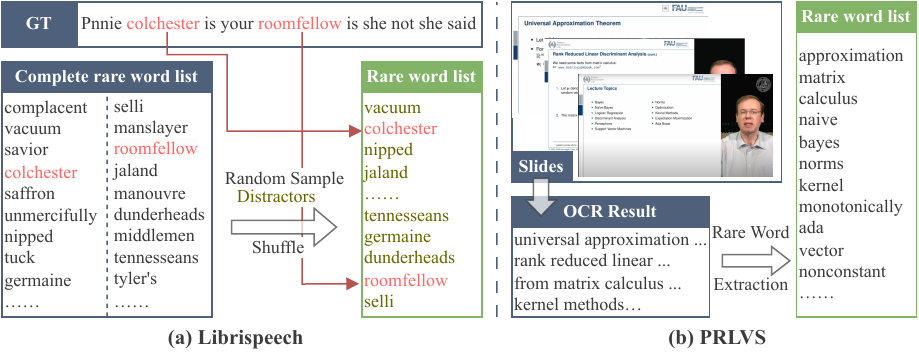}
  \vspace{-3mm}
  \caption{Process for creating the rare word list: (a) Simulating real-world tasks using the LibriSpeech dataset, following the method described in \cite{le2021contextualized}, which is also applied to the ATIS and SNIPS datasets. (b) Extracting new words from slides for experiments in real-world scenarios.}
  \label{fig:rw_list}
  \vspace{-5mm}
\end{figure*}

\vspace{-3mm}
\subsection{Datasets}
To evaluate the effectiveness and robustness of our proposed PMF-CEC, we conducted experiments on five datasets generated using different ASR engines. The dataset statistics are summarized in Table~\ref{tab:dataset_utterances}.


\begin{itemize}[leftmargin=*]
\setlength{\itemsep}{0pt}
\setlength{\parsep}{0pt}
\setlength{\parskip}{0pt}
\item \textbf{ATIS} \cite{sundararaman2021phoneme} included 8 hours of audio recordings of users making flight reservations, along with manually transcripts. The ASR transcripts were generated by a \texttt{LAS} system \cite{chan2016listen}.

\item \textbf{SNIPS} \cite{9054689} was designed for natural language understanding and was collected from the voice assistant. The ASR transcripts were generated using the \texttt{Kaldi} toolkit \cite{povey2011kaldi}\footnote{\url{https://github.com/kaldi-asr/kaldi}}.

\item \textbf{Librispeech} \cite{2015Librispeech} was a collection of 960 hours of audiobooks. The \texttt{Wenet} \cite{yao2021wenet} ASR toolkit\footnote{\url{https://github.com/wenet-e2e/wenet/tree/main}} was utilized to obtain the related ASR transcripts. The \textit{dev-clean} and \textit{dev-other} datasets were used for validation, whereas the \textit{test-clean} and \textit{test-other} datasets were used for evaluation. 

\item \textbf{DATA2} \cite{yadav2020end} was designed for E2E spoken named entity recognition (NER) tasks, containing 70,763 speech-to-text pairs with annotated entity words or phrases. We randomly split the data into 2,000 validation pairs, 2,000 test pairs, and the remainder for training. We utilized the \texttt{ESPNet} \cite{watanabe2018espnet} ASR toolkit\footnote{\url{https://github.com/espnet/espnet/tree/master}} to obtain the corresponding ASR transcripts. 


\item \textbf{PRLVS} \cite{hernandez2021multimodal} comprised a full-semester pattern recognition course, including 43 lecture videos with accompanying slides, totaling 11.4 hours of content. We employed the \texttt{SpeechBrain} \cite{ravanelli2021speechbrain}  ASR toolkit\footnote{\url{https://github.com/speechbrain/speechbrain}} to obtain the transcripts.
\vspace{-3mm}
\end{itemize}

\begin{table}[t]
\centering
\caption{Numbers of utterances in different datasets.}
\vspace{-3mm}
\label{tab:dataset_utterances}
\begin{tabular}{l|cccccc}
\toprule
\multirow{2}{*}{\textbf{Dataset}} & \multirow{2}{*}{\textbf{ATIS}} & \multirow{2}{*}{\textbf{SNIPS}} & \multicolumn{2}{c}{\textbf{Librispeech}} & \multirow{2}{*}{\textbf{DATA2}} & \multirow{2}{*}{\textbf{PRLVS}} \\
 & & & \textbf{Clean} & \textbf{Other} & & \\
\midrule
\midrule
\textbf{Train} & 3,867 & 13,084 & 132,553 & 148,688 & 66,763 & 3,680 \\
\textbf{Valid} & 967 & 700 & 2,703 & 2,864 & 2,000 & 460 \\
\textbf{Test} & 800 & 700 & 2,620 & 2,939 & 2,000 & 460 \\
\bottomrule
\end{tabular}
\end{table}
%


\subsection{Rare Word List Construction}
\label{sec:rw_con}

Owing to the lack of available rare word lists for the ATIS, SNIPS, and Librispeech datasets, we followed the validated simulation proposed in \cite{le2021contextualized} to construct rare word lists for each dataset, as shown in Fig. \ref{fig:rw_list}(a). Specifically,  a complete rare word list was initially compiled for the Librispeech dataset, consisting of 209.2K distinct words, by excluding the top 5,000 most common words from the Librispeech LM training corpus. Rare words are defined as words that belong to the complete rare word list in each dataset. 
Next, the rare word lists were constructed by identifying words from the reference of each utterance that were present in the complete rare word list. Additionally, a specified number of distractors (e.g., 1,000) were added to each rare word list, as determined by the experiment requirements. By utilizing this approach, we can effectively organize the rare word lists for each utterance\footnote{\url{https://github.com/facebookresearch/fbai-speech/tree/master/is21_deep_bias}}, containing words from the complete rare word list and supplementing them with distractors. Similar methods were employed to construct rare word lists for the remaining ATIS and SNIPS datasets.

\begin{table}[t]
\centering
\caption{Coverage of rare word lists on the evaluation sets (\%). Coverage is the total number of rare words divided by the total number of words in each set.}
\vspace{-3mm}
\label{tab:coverage}
\begin{tabular}{l|cccccc}
\toprule
\multirow{2}{*}{\textbf{Dataset}} & \multirow{2}{*}{\textbf{ATIS}} & \multirow{2}{*}{\textbf{SNIPS}} & \multicolumn{2}{c}{\textbf{Librispeech}} & \multirow{2}{*}{\textbf{DATA2}} & \multirow{2}{*}{\textbf{PRLVS}} \\
 & & & \textbf{Clean} & \textbf{Other} & & \\
\midrule
\midrule
\textbf{Coverage} & 27.08 & 18.11 & 10.73 & 9.87 & 5.13 & 8.42 \\
\bottomrule
\end{tabular}
\vspace{-3mm}
\end{table}

To demonstrate the feasibility of obtaining rare word lists in practice, we focused on the PRLVS dataset. The construction process involved collecting slides for each lecture and utilizing the Tesseract 4 OCR engine\footnote{\url{https://github.com/tesseract-ocr/tesseract}} for text extraction, as shown in Fig. \ref{fig:rw_list}(b). Distinct word tokens were then extracted from the OCR output files. Among these tokens, only those belonging to the complete rare word list or appearing fewer than 15 times in the PRLVS train set were included in the lecture-specific rare word list. These rare word lists were subsequently applied for the context-aware correction of all utterances within the corresponding lectures \cite{sun2022tree}.

Additionally, to verify the extensibility of our PMF-CEC method to phrases, we selected the DATA2 dataset, which was used for spoken NER tasks. Since this dataset contains labels related to words or phrase entities such as persons, organizations, and locations, we treated these entities as rare words.
The proportion of rare words in each test set is shown in Table \ref{tab:coverage}.

\subsection{Evaluation Metrics}
We employed the following four evaluation metrics to evaluate the performance:

\begin{table*}[ht]
\centering
\caption{Measurements of error correction performance on five datasets (\%).}
\vspace{-3mm}
\label{tab:error_correction}
\resizebox{\linewidth}{!}{
\begin{tabular}{l|cccccc}
\toprule

\multirow{4}{*}{\textbf{Method}} & \multirow{2}{*}{\textbf{ATIS}} & \multirow{2}{*}{\textbf{SNIPS}} & \multicolumn{2}{c}{\textbf{Librispeech}} & \multirow{2}{*}{\textbf{DATA2}} & \multirow{2}{*}{\textbf{PRLVS}} \\
 & & & \textbf{Test-clean} & \textbf{Test-other} & & \\

\cmidrule{2-7} 
     & WER/WERR & WER/WERR & WER/WERR & WER/WERR & WER/WERR & WER/WERR\\
      & (U-WER/B-WER) & (U-WER/B-WER) & (U-WER/B-WER) & (U-WER/B-WER) & (U-WER/B-WER) & (U-WER/B-WER)\\

\midrule
\midrule
\multirow{2}{*}{Original} & 30.65/- & 45.73/- & 3.51/- & 9.57/- & 7.94/- & 18.66/- \\
 & (20.58/87.78) & (34.20/99.64) & (1.07/13.31) & (3.22/30.63) & (6.90/26.52) & (10.66/47.24) \\
\midrule
\multirow{2}{*}{SC\_BART \cite{zhao2021bart}} & 21.47/29.95 & 30.35/33.63 & 3.12/11.11 & 9.05/5.43 &7.23/8.94 & 15.14/18.86 \\
 & (14.63/49.25) & (21.86/70.25) & (1.20/12.15) & (3.76/26.83) & (6.18/25.18) & (10.43/27.67) \\
\midrule
\multirow{2}{*}{distillBART \cite{shleifer2020pre}} & 26.51/13.51 & 33.28/27.23 & 3.35/4.56 & 9.37/2.40 & 7.62/4.03 & 17.98/3.64 \\
 & (18.54/74.67) & (24.08/76.43) & (1.26/12.75) & (4.42/27.11) &(6.54/26.36) & (10.64/43.57) \\
\midrule
\multirow{2}{*}{ConstDecoder$_{\text{trans}}$ \cite{DBLP:journals/corr/abs-2208-04641}} & 21.74/29.07 & 30.98/32.25 & 3.27/6.84 & 9.35/2.30 & 7.41/6.68 & 15.31/17.95 \\
 & (14.78/50.57) & (22.09/71.43) & (1.22/12.47) & (4.38/27.67) & (6.11/27.20) & (10.95/28.33) \\
\midrule
\midrule
& 18.95/38.17 & 28.57/37.52 & 2.71/22.80 & 8.23/14.00 & 5.39/32.16 & 13.17/29.42 \\
 \multirow{-2}{*}{ED-CEC \cite{he2023ed}} & (14.88/38.38) & (21.47/62.79) & (1.25/9.72) & 
 (3.85/21.02) &
 (4.72/15.11) & (10.01/20.72) \\
 \midrule
\rowcolor[HTML]{EFEFEF} 
& \textbf{16.21/47.11} & \textbf{25.34/44.59} & \textbf{2.52/28.21} & \textbf{7.88/17.66} & \textbf{5.11/35.64} & \textbf{11.68/37.41} \\
\rowcolor[HTML]{EFEFEF} 
 \multirow{-2}{*}{\textbf{PMF-CEC (Proposed)}} & \textbf{(13.25/35.16)} & \textbf{(19.89/56.37)} & \textbf{(1.12/8.54)} & 
 \textbf{(3.35/18.38)} &
 \textbf{(4.69/13.69)} & \textbf{(9.31/18.56)} \\
\bottomrule
\end{tabular}
}
\vspace{-3mm}
\end{table*}

\begin{table}[ht]
\centering
\caption{Average inference time in ms.}
\vspace{-3mm}
\label{tab:inference_speed}
\resizebox{1\linewidth}{!}{
\begin{tabular}{lccccc}
\toprule
\textbf{Method} & \textbf{ATIS} & \textbf{SNIPS} & \textbf{Librispeech} & \textbf{DATA2} & \textbf{PRLVS} \\
\midrule
\midrule
SC\_BART \cite{zhao2021bart} & 90.30 & 75.30 & 144.32 & 156.12 & 103.62 \\
distillBART \cite{shleifer2020pre} & 45.55 & 41.55 & 69.80 & 75.74 & 57.60 \\
ConstDecoder$_{\text{trans}} \cite{DBLP:journals/corr/abs-2208-04641}$ & 25.61 & 26.66 & 23.69 & 21.39 & 18.29 \\
ED-CEC \cite{he2023ed} & 32.59 & 31.87 & 30.16 & 24.93 & 22.86 \\
\midrule
\rowcolor[HTML]{EFEFEF} 
PMF-CEC (Proposed) & 38.12 & 36.94 & 34.48 & 29.19 & 27.21 \\
vs SC\_BART & 2.8$\times$ & 2.4$\times$ & 4.2$\times$ & 5.3$\times$ & 4.5$\times$ \\
vs distillBART & 1.4$\times$ & 1.3$\times$ & 2.0$\times$ & 2.6$\times$ & 2.5$\times$ \\
vs ConstDecoder$_{\text{trans}}$ & 0.8$\times$ & 0.8$\times$ & 0.7$\times$ & 0.7$\times$ & 0.8$\times$ \\
vs ED-CEC & 0.9$\times$ & 0.9$\times$ & 0.9$\times$ & 0.9$\times$ & 0.8$\times$ \\
\bottomrule
\end{tabular}}
\vspace{-3mm}
\end{table}

\begin{itemize}[leftmargin=*]
\setlength{\itemsep}{0pt}
\setlength{\parsep}{0pt}
\setlength{\parskip}{0pt}
\item \textbf{WER} is the overall word error rate on all words.
\item \textbf{WERR} quantifies the WER reduction across all words.
\item \textbf{U-WER} calculates the unbiased WER on words not included in the rare word list.
\item \textbf{B-WER} computes the biased WER on words present in the rare word list.
\item \textbf{RW-Recall} measures the recall of the rare words that are correctly identified in the utterance.
\end{itemize}

 In the case of insertion errors, if the inserted word is found in the rare word list, it will contribute to B-WER; otherwise, it will be considered for U-WER. The objective of contextualization is to improve B-WER while minimizing any significant degradation in U-WER \cite{le2021contextualized}.

\section{Experimental Results}
\label{sec:results}

\subsection{Comparisons of Baseline AEC Methods}
In the experiments, we compared five baseline AEC methods with our proposed PMF-CEC. The comparison was conducted using the following methods across five public datasets:
\begin{itemize}[leftmargin=*]
\setlength{\itemsep}{0pt}
\setlength{\parsep}{0pt}
\setlength{\parskip}{0pt}
\item \textbf{Original} denotes the original ASR transcripts.
\item \textbf{SC\_BART} \cite{zhao2021bart} has demonstrated superior performance in AEC tasks, achieving SOTA results.
\item \textbf{distillBART} \cite{shleifer2020pre} is a distilled version of the BART large model.
\item \textbf{ConstDecoder$_{\textbf{\textit{trans}}}$} \cite{DBLP:journals/corr/abs-2208-04641} is a constrained decoding method designed to improve the inference speed of AEC while preserving a certain level of error correction performance.

\begin{table*}[h!]
    \centering
    \caption{WER and B-WER on LibriSpeech test sets, PRLVS test set and DATA2 test set using Whisper and TCPGen, rescored with GPT-2 (\%), and corrected by PMF-CEC. 
    LM weights are finetuned on each validation set separately.}
    \vspace{-3mm}
    \label{tab:whisper_result}
    \begin{tabular}{l|cccccccc}
        \toprule
        \multirow{2}{*}{\textbf{Method}} & \multicolumn{2}{c}{\textbf{Librispeech Test-clean}} & \multicolumn{2}{c}{\textbf{Librispeech Test-other}} & \multicolumn{2}{c}{\textbf{DATA2}} & \multicolumn{2}{c}{\textbf{PRLVS}} \\
        \cmidrule(lr){2-3} \cmidrule(lr){4-5} \cmidrule(lr){6-7} \cmidrule(lr){8-9}
         & WER $\downarrow$ & B-WER $\downarrow$ & WER $\downarrow$ & B-WER $\downarrow$ & WER $\downarrow$ & B-WER $\downarrow$ & WER $\downarrow$ & B-WER $\downarrow$ \\
        \midrule
        \midrule
        Whisper meduim.en & 4.00 & 15.08 & 6.83 & 21.89 & 4.96 & 15.22 & 6.74 & 19.83 \\
         \hspace{1em} + TCPGen \cite{sun2021tree} & 3.46 & 12.05 & 6.41 & 18.15 & 4.02 & 11.64 & 6.27 & 15.77 \\
         \hspace{1em} + GPT-2 \cite{radford2019language} & 3.88 & 14.91 & 6.74 & 21.34 & 4.51 & 13.98 & 6.50 & 18.26 \\
         \hspace{1em} + PMF-CEC & 2.99 & 9.61 & 6.02 & 16.45 & 3.31 & 8.82 & 5.99 & 12.51 \\
         \rowcolor[HTML]{EFEFEF} 
         \hspace{1em} \textbf{+ TCPGen + GPT-2 + PMF-CEC} & \textbf{2.57} & \textbf{8.17} & \textbf{5.85} & \textbf{15.50} & \textbf{3.04} & \textbf{7.27} & \textbf{5.56} & \textbf{10.67} \\
        \bottomrule
    \end{tabular}
    \vspace{-3mm}
\end{table*}



\item \textbf{ED-CEC} \cite{he2023ed}, our previous work, introduces a contextual ASR postprocessing method that enhances the recognition of rare words by detecting errors and using context-aware correction, optimizing the decoding process and leveraging a rare word list for improved accuracy.
\end{itemize}

Table \ref{tab:error_correction} shows that when the rare word list size is set to 100, our model significantly improves WER results across all five public datasets compared with previous AEC baseline models. Specifically, our model PMF-CEC achieves substantial reductions in B-WER compared with baseline AEC methods without contextual biasing (i.e., Original, SC\_BART, distillBART, and ${\rm ConstDecoder}_{trans}$). Notably, compared with the SOTA model SC\_BART, our model demonstrates a relative B-WER improvement ranging from 19.76\% to 45.63\%, highlighting its effectiveness in correcting rare words. Furthermore, compared with our previous ED-CEC, the introduction of the phoneme-augmented multimodal fusion method and RPM further enhances the error correction capability and reduces overcorrection, achieving average relative improvements of 8.22\% in U-WER and 10.52\% in B-WER.

Table \ref{tab:inference_speed} denotes that our method achieves a 2.4 to 5.3 times improvement in inference speed compared with the autoregressive SOTA baseline model SC\_BART. Although the inference speed is slightly lower than those of the partially autoregressive baseline models ${\rm ConstDecoder}_{trans}$ and ED-CEC, the error correction performance is significantly improved, demonstrating that our model strikes a good balance between WERR and inference speed. Additionally, the performance improvements across five different prominent ASR systems demonstrate the robustness of our method.

\vspace{-3mm}
\subsection{Comparisons with Other Contextual Biasing Methods}
To validate the effectiveness of our PMF-ECE on the large-scale ASR model \texttt{Whisper} \cite{radford2022robust, fu2023robust}, we selected the general dataset Librispeech and two domain-specific datasets, PRLVS and DATA2. We applied our proposed method to the ASR transcriptions obtained by \texttt{whisper-medium.en}\footnote{\url{https://huggingface.co/openai/whisper-medium.en}}. Additionally, to compare with other contextual biasing methods, we incorporated the following two mainstream approaches:

\begin{itemize}[leftmargin=*]
\setlength{\itemsep}{0pt}
\setlength{\parsep}{0pt}
\setlength{\parskip}{0pt}
\item \textbf{TCPGen} \cite{sun2021tree} integrates a rare word list into E2E ASR models by structuring the words into an efficient prefix tree and creating a neural shortcut to enhance the recognition of these words during decoding. During training, we kept the \texttt{Whisper} parameters frozen and only trained the TCPGen component.

\item \textbf{GPT-2} \cite{radford2019language} is a large pretrained LM based on the Transformer architecture, trained to predict the next token through self-supervised learning. The training data comprises text from a social media platform, amounting to over ten billion words\footnote{\url{https://openai.com/research/gpt-2-1-5b-release}}. 
We utilized \texttt{GPT-2}\footnote{\url{https://huggingface.co/openai-community/GPT-2}} without finetuning to rescore the 50-best hypotheses list, with the weight of the LM optimized for each validation set.
\end{itemize}

Notably, to ensure comparability and compatibility with other contextual biasing setups \cite{sun2021tree, sun2022tree, sun2024graph, le2021contextualized}, text normalization was not applied during evaluation unless otherwise specified. A detailed discussion on this can be found in Section \ref{sec:discussion}.

As shown in Table \ref{tab:whisper_result}, our method remains effective for ASR transcriptions obtained from large-scale models when the size of the rare word list size is set to 100. Specifically, our PMF-CEC achieves relative B-WER reductions of 36.27\% and 24.85\% on test-clean and test-other sets of Librispeech, respectively, a relative R-WER reduction of 42.05\% on DATA2, and a relative B-WER reduction of 36.91\% on PRLVS. The reduction in B-WER is more pronounced on domain-specific datasets than on the generic LibriSpeech dataset, because the rare word lists during the LibriSpeech testing stage mainly consist of generic words, whereas those for the other two test sets include domain-specific terms such as names, locations, or professional terms.

Additionally, the TCPGen module also significantly reduces B-WER, whereas GPT-2, serving as a general knowledge source, does not exhibit significant improvements in reducing B-WER. We observed that our PMF-CEC reaches the average relative B-WER reductions of 18.63\% and 31.72\% compared with TCPGen and GPT-2, respectively, highlighting the advantages of our contextual biasing approach. Finally, combining all three methods further reduces WER, achieving optimal performance.

    

    

\subsection{Comparisons with LLM-based ASR and AEC Methods}

To further evaluate the effectiveness of PMF-CEC, we compare it against three representative LLM-based ASR and AEC methods on the DATA2 dataset:

\begin{itemize}[leftmargin=*]
\setlength{\itemsep}{0pt}
\setlength{\parsep}{0pt}
\setlength{\parskip}{0pt}


\item \textbf{Whispering LLaMA ($\mathcal{WL}$)} \cite{radhakrishnan2023whispering} is a cross-modal generative AEC model that integrates audio and text features for instruction-guided ASR correction. It is initialized from \texttt{Alpaca}\footnote{\url{https://github.com/tatsu-lab/stanford_alpaca}}, a model finetuned from \texttt{LLaMA-7B} \cite{touvron2023llama}. 
To enable a fair comparison with $\mathcal{WL}$, we adopt the same experimental setup: the 5-best ASR hypotheses are generated using \texttt{whisper-tiny}\footnote{\url{https://huggingface.co/openai/whisper-tiny}}, and audio features are extracted from \texttt{whisper-large-v2}\footnote{\url{https://huggingface.co/openai/whisper-large-v2}}. For $\mathcal{WL}$, we use the medium variant $\mathcal{WL}_M$ with a LoRA rank of $r=16$.


\item \textbf{SLAM-ASR} \cite{ma2024embarrassingly} is an LLM-based E2E ASR framework that connects a speech encoder and a large language model via a linear projection layer. In our experiments, we use \texttt{WavLM-Large}\footnote{\url{https://huggingface.co/microsoft/wavlm-large}} as the speech encoder and \texttt{Vicuna-7B}\footnote{\url{https://huggingface.co/lmsys/vicuna-7b-v1.5}} as the LLM backbone, finetuned with LoRA adaptation (rank $r=16$). The encoder output is downsampled and linearly projected to match the LLM input dimension.




\item \textbf{MaLa-ASR} \cite{yang24f_interspeech} builds upon SLAM-ASR by introducing rare-word biasing via domain-specific keyword prompting. We follow the same model architecture as SLAM-ASR. Keywords are prepended to the prompt to improve recognition of rare or domain-specific terms.
\end{itemize}

As shown in Table \ref{tab:performance_llm_aed}, $\mathcal{WL}_M$ performs generative AEC over 5-best hypotheses and achieves a lower WER of 15.61\% compared with 20.90\% for PMF-CEC applied to the Whisper 1-best output. However, PMF-CEC yields a significantly lower B-WER, demonstrating stronger bias correction capability. Notably, combining the two approaches ``$\mathcal{WL}_M$ + PMF-CEC" further improves performance.

Table \ref{tab:performance_llm_asr} further compares PMF-CEC with SLAM-ASR and MaLa-ASR. When integrated into SLAM-ASR with 100 biasing words, PMF-CEC reduces WER from 5.75\% to 4.66\% and B-WER from 23.44\% to 16.13\%, with only a 0.04-second increase in inference time. The performance remains stable as the biasing list scales up, confirming the robustness and scalability of our model and its applicability to LLM-based ASR systems.
In contrast, MaLa-ASR achieves a lower B-WER than SLAM-ASR when using a small biasing list of 100, with B-WER reduced to 5.75\% compared with 16.13\%. However, its WER increases to 5.87\%, suggesting that biasing word prompts may interfere with overall semantic modeling. When the number of bias terms increases to 1000, MaLa-ASR experiences severe performance degradation, with WER reaching 116.99\% and B-WER reaching 100.00\%, revealing poor robustness under large-scale prompting conditions.

Overall, thanks to its lightweight design, PMF-CEC offers superior efficiency and more reliable correction performance than LLM-based methods, making it well-suited for latency-sensitive, real-world ASR applications.

\begin{table}[t]
\centering
\caption{WER and B-WER comparison of the LLM-based AEC model $\mathcal{WL}_M$ and our PMF-CEC model on the DATA2 test set (\%). We use Whisper Tiny to generate 5-best hypotheses. ``Oracle'' refers to the candidate with the lowest WER compared with the ground truth within the 5-best hypotheses. The unit of inference time is seconds per sentence (s/sentence).}
\vspace{-3mm}
\label{tab:performance_llm_aed}
\begin{adjustbox}{max width=\columnwidth}
\begin{tabular}{l|c|c|c}
\toprule
\textbf{Model} & \textbf{Biasing Size} & \textbf{WER}$\downarrow$ / \textbf{B-WER}$\downarrow$ & \textbf{Inference Time} \\
\midrule
\midrule
Whisper Oracle & - & 13.87 / 45.48 & - \\
$\mathcal{WL}_M$ \cite{radhakrishnan2023whispering} & - & 15.61 / 44.73 & 3.47\\
Whisper 1-best & - & 25.69 / 56.51 & - \\

\hspace{1em} + PMF-CEC & 100 & 20.90 / 30.82 & \textbf{0.10} \\
\rowcolor[HTML]{EFEFEF} 
$\mathcal{WL}_M$ \cite{radhakrishnan2023whispering} + PMF-CEC & 100 & \textbf{13.14} / \textbf{28.33} & 3.58\\
\bottomrule
\end{tabular}
\end{adjustbox}
\end{table}

\begin{table}[t]
\centering
\caption{WER and B-WER comparison of different LLM-based ASR models and our PMF-CEC model on the DATA2 test set (\%). The unit of inference time is seconds per sentence (s/sentence).}
\vspace{-3mm}
\label{tab:performance_llm_asr}
\begin{adjustbox}{max width=\columnwidth}
\begin{tabular}{l|c|c|c}
\toprule
\textbf{Model} & \textbf{Biasing Size} & \textbf{WER}$\downarrow$ / \textbf{B-WER}$\downarrow$ & \textbf{Inference Time} \\
\midrule
\midrule
SLAM-ASR \cite{ma2024embarrassingly} & - & 5.75 / 23.44 & 2.34\\
\rowcolor[HTML]{EFEFEF} 
\hspace{1em} + PMF-CEC & 100 & \textbf{4.66} / 16.13 & \textbf{+ 0.04} \\
& 300 & 4.69 / 16.36 & + 0.13\\
& 500 & 4.72 / 16.45 & + 0.34\\
& 1000 & 4.78 / 16.71 & + 0.53\\
\midrule
\rowcolor[HTML]{EFEFEF} 
MaLa-ASR \cite{yang24f_interspeech} & 100 & 5.87 / \textbf{5.75} & 3.52\\
& 300 & 10.03 / 12.59 & 5.92\\
& 500 & 31.61 / 44.02 & 7.93\\
& 1000 & 116.99 / 100.00 & 10.57\\
\bottomrule
\end{tabular}
\end{adjustbox}
\vspace{-3mm}
\end{table}


\begin{table}[t]
    \centering
    \vspace{-3mm}
    \caption{Impact of each module in PMF-CEC evaluated via ablation experiments on Librispeech test-clean set (\%).  Accuracy refers to the AED module’s detection accuracy.}
    \vspace{-2mm}
    \label{tab:ablation_study}
    \begin{tabular}{l|ccc}
        \toprule
      \textbf{Method}  & \textbf{WER $\downarrow$} & \textbf{B-WER $\downarrow$} & \textbf{AED Accuracy $\uparrow$}\\
        \midrule
        \midrule
        \rowcolor[HTML]{EFEFEF}
        \textbf{PMF-CEC (full)} & \textbf{2.52} & \textbf{8.54} & \textbf{97.40}\\
        \hspace{1em} w/o Phoneme Encoder & 2.62 & 9.55 & 97.40\\
        \hspace{1em} w/o Context Decoder & 3.18 & 12.13 & 97.40\\
        \hspace{1em} w/o RPM & 2.61 & 8.78 & 95.73 \\
        \bottomrule
    \end{tabular}
    \vspace{-3mm}
    
\end{table}

\vspace{-3mm}
\subsection{Impact of Individual Modules in PMF-CEC}

To evaluate the individual contributions of each component in the PMF-CEC model, we conduct ablation studies on the LibriSpeech test-clean dataset, as shown in Table \ref{tab:ablation_study}.

First, removing the phoneme encoder increased the WER from 2.52\% to 2.62\%, and the B-WER from 8.54\% to 9.55\%. This indicates that phoneme-level representations help the model make more accurate decisions in the presence of acoustic ambiguity or homophonic interference, particularly when correcting biased homophone substitutions.

Second, the removal of the context decoder resulted in the most significant performance degradation, with WER rising to 3.18\% and B-WER to 12.13\%. This demonstrates that the context decoder effectively integrates linguistic context and rare-word biasing information, allowing the model to prioritize contextually appropriate biasing terms among multiple candidates and thereby significantly improve biasing correction accuracy.

Finally, removing the RPM increased the WER to 2.61\%, while the accuracy of editing operations in the AED module rose from 95.73\% to 97.40\%. These results suggest that RMP plays a crucial role in filtering out low-confidence edits, preventing unnecessary corrections, and thus mitigating overcorrection to enhance the overall stability and quality of the model outputs.

\vspace{-4mm}
\subsection{Impact of Rare Word List Size}

\begin{figure}[htbp]

\begin{minipage}[b]{1\linewidth}
  \centering
  \centerline{\includegraphics[width=8cm]{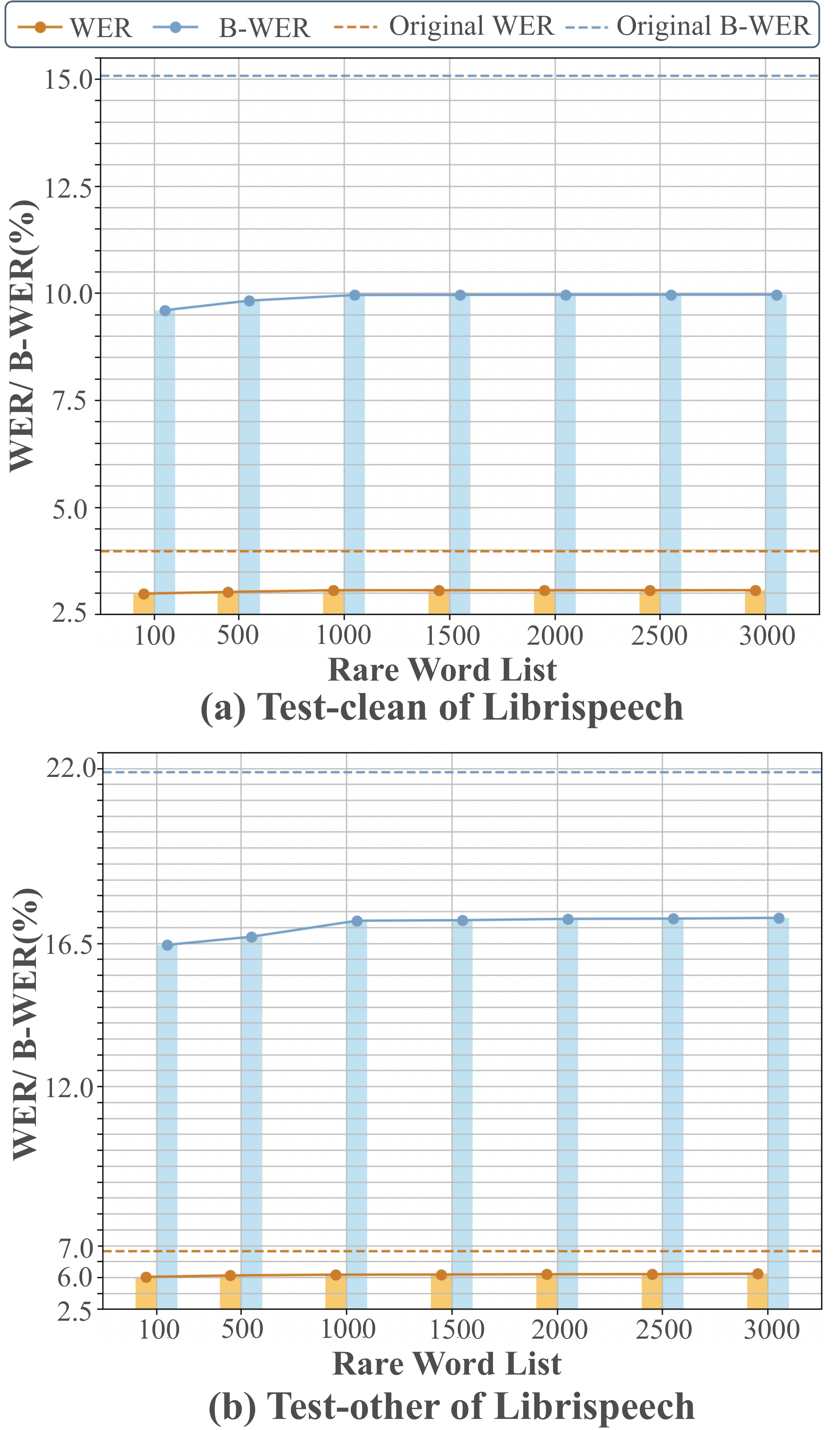}}
\end{minipage}
\vspace{-6mm}
\caption{WER results of Librispeech test sets with varying rare word list sizes.}
\vspace{-6mm}
\label{fig:size}
\end{figure}

To investigate how the rare word list size affects the performance of PMF-CEC, we conducted experiments on the LibriSpeech test-clean and test-other datasets, based on the results shown in Table~\ref{tab:whisper_result}. Rare word lists of varying sizes (from 100 to 3000 words) were extended with distractors. As shown in Fig.~\ref{fig:size}, the model achieves the lowest WER and B-WER on both test sets when the rare word list size is 100.

Increasing the rare word list size to 3000 shows a slight upward trend in WER and B-WER; nonetheless, relative reductions of 34.81\% and 20.92\% in B-WER compared with the original ASR transcripts demonstrated the robustness of PMF-CEC against large rare word lists. It is noteworthy that in the case where the rare word list is empty (take the test-clean set as an example), although WER significantly increases to 3.69\%, PMF-CEC still manages to partially correct errors by generating decoder outputs, achieving a 7.75\% relative WERR compared with the original ASR transcripts. This underscores our method's capability to generate relevant words or select them from rare word lists. However, correcting for rare words still requires a rare word list containing relevant words.

Additionally, an ``anti-context" experiment was conducted where 100 irrelevant distractor words were chosen as the rare word list. Similarly, in the test-clean set scenario, this resulted in a WER of 3.72\%, achieving a 7.00\% relative WERR compared with the original ASR transcripts. This result demonstrated that PMF-CEC effectively avoids selecting irrelevant context words when lacking relevant context, instead correcting errors by generating partially correct words through the generation decoder.

\begin{figure*}[!t]
  \centering
  \hspace*{-0.1cm}  
  \includegraphics[width=2.05\columnwidth]{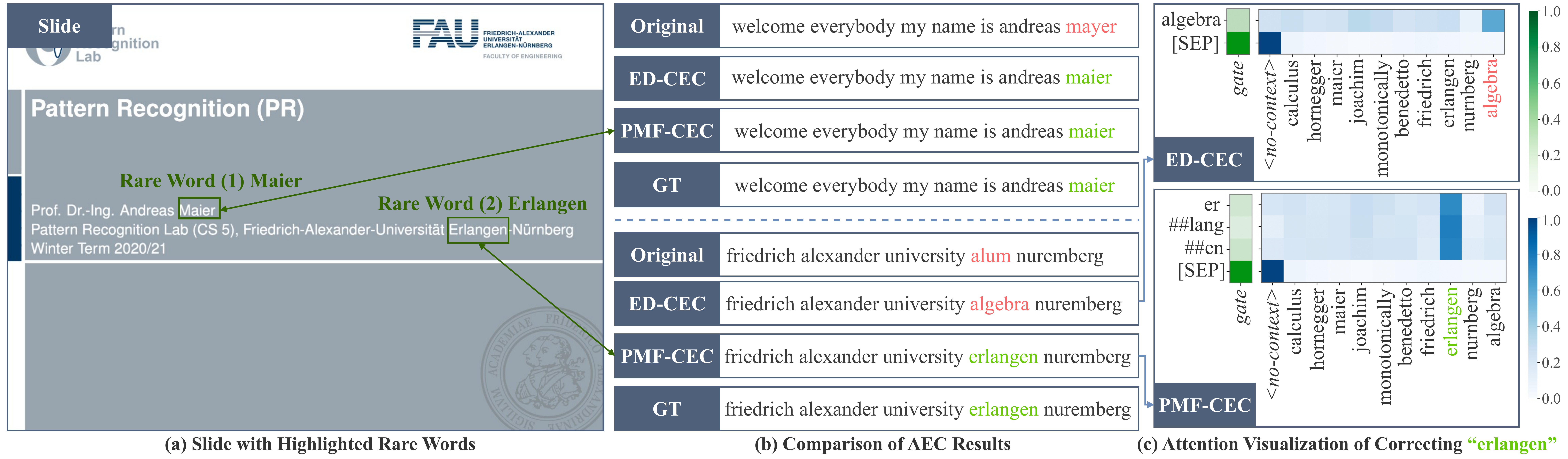}
  \vspace{-8mm}
  \caption{Examples of correcting the rare words ``maier" and ``erlangen" in the PRLVS dataset. We present slides with the rare words, the original ASR transcription, the corrected results using the ED-CEC method, the corrected results using our PMF-CEC method, and the ground truth (GT) transcription. Errors are marked in red, and correctly corrected parts are highlighted in green. Additionally, we provide heatmaps illustrating the correction process for the rare word "erlangen" using both methods.}
  \label{fig:example_visual}
  \vspace{-2mm}
\end{figure*}

\begin{figure}[!t]
  \centering
\vspace{-8mm}
  \hspace*{-0.5cm}  
  \includegraphics[width=1.1\columnwidth]{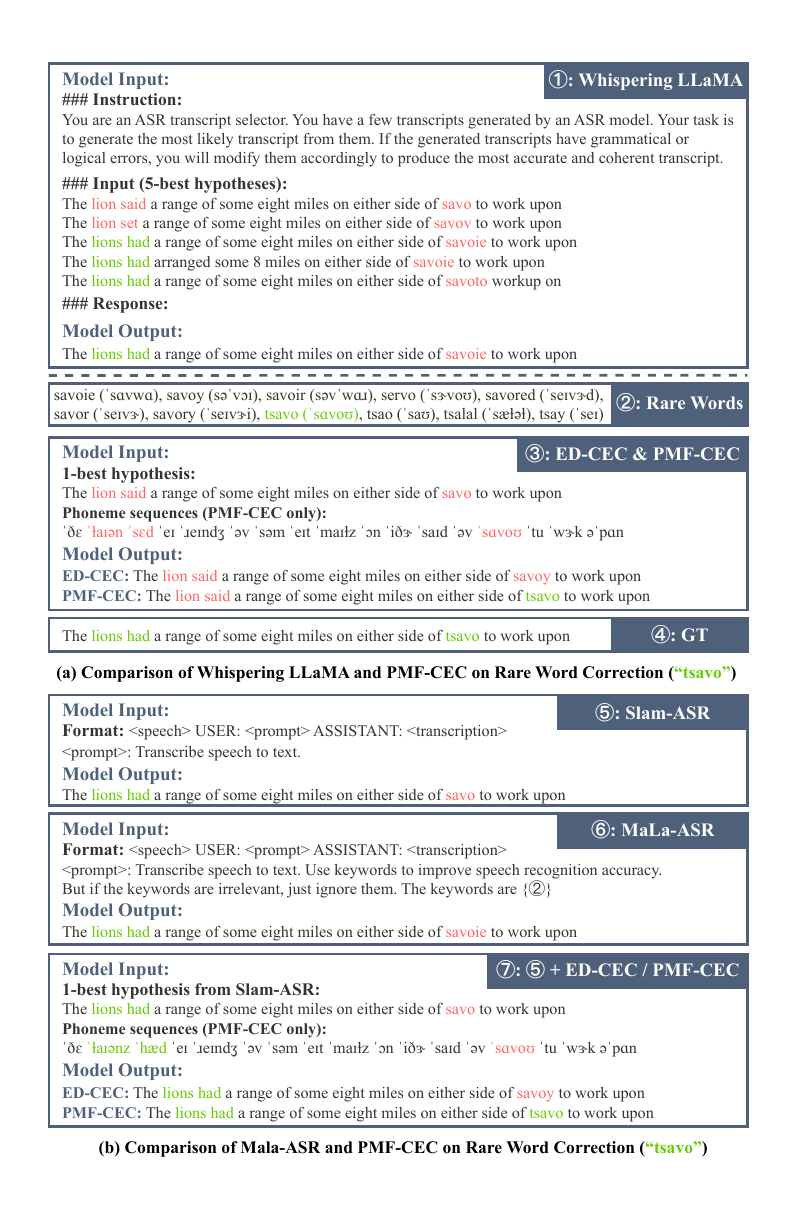}
  \vspace{-16mm}
  \caption{An example from the DATA2 test set illustrating the challenge of rare word recognition under homophone confusion. Red indicates errors; green highlights correct recovery.}
  \label{fig:example_llm}
  \vspace{-7mm}
\end{figure}

\begin{table*}[ht]
\centering
\caption{Analysis of few-shot generalization ability on the Librispeech test-clean and test-other and PRLVS test set. The results show B-WER $\downarrow$ / RW-Recall $\uparrow$ (\%).}
\vspace{-3mm}
\label{tab:few_shot}
\begin{tabular}{l|ccccccccc}
\toprule
\multirow{2}{*}{\textbf{Model}} & \multicolumn{3}{c}{ \textbf{Librispeech Test-clean}} & \multicolumn{3}{c}{\textbf{Librispeech Test-other}} & \multicolumn{3}{c}{\textbf{PRLVS Test Set}}\\
\cmidrule(lr){2-4} \cmidrule(lr){5-7} \cmidrule(lr){8-10}
 & $\leq$ 0-shot & $\leq$ 5-shot & $\leq$ 100-shot & $\leq$ 0-shot & $\leq$ 5-shot & $\leq$ 100-shot & $\leq$ 0-shot & $\leq$ 5-shot & $\leq$ 100-shot\\
\midrule
\midrule
{Original} & 50.91/46.05 & 32.94/65.66 & 13.43/87.21 & 67.30/32.58 & 51.94/45.66 & 25.22/73.46 & 64.83/40.58 & 41.51/60.16 & 17.32/80.34 \\
ED-CEC & 38.43/62.09 & 25.82/73.14 & 7.87/92.78 & 57.63/46.30 & 44.92/57.12 & 20.03/79.57 & 56.10/49.33 & 30.12/69.98 & 13.01/86.73\\
\rowcolor[HTML]{EFEFEF} 
\textbf{PMF-CEC} & 
\textbf{35.72/69.11} & \textbf{23.37/78.26} & \textbf{7.52/95.75} & \textbf{56.13/48.31} & \textbf{43.84/58.45} & \textbf{19.57/80.84} & \textbf{55.79/50.62} & \textbf{27.83/72.55} & \textbf{12.42/88.25}\\
\bottomrule
\end{tabular}
\vspace{-3mm}
\end{table*}

\begin{table}[ht]
    \centering
    \caption{Results of domain adaptation with limited data on PRLVS and DATA 2 (\%).}
    \vspace{-3mm}
    \label{tab:domain}
    \begin{threeparttable} 
    \begin{tabular}{l|cccc}
        \toprule
        \multirow{2}{*}{\textbf{Method}} & \multicolumn{2}{c}{\textbf{PRLVS}} & \multicolumn{2}{c}{\textbf{DATA2}} \\
        \cmidrule(lr){2-3} \cmidrule(lr){4-5}
        & WER ↓ & B-WER ↓ & WER ↓ & B-WER ↓ \\
        \midrule
        \midrule
        Original                   & 6.74 & 19.83 & 4.96 & 15.22 \\
         PT  & 6.89 & 18.33 & 5.12 & 13.78 \\
        FT (Full Data) \tnote{$\diamond$}                           & 5.99 & 12.51 & 3.31 & 8.82 \\
        PT + FT (10\% Data)                          & 6.29 & 14.79 & 3.17 & 7.63 \\
        \rowcolor[HTML]{EFEFEF} 
        \textbf{PT + FT (Full Data)}                            & \textbf{5.40} & \textbf{10.27} & \textbf{2.72} & \textbf{6.35} \\

        \bottomrule
    \end{tabular}
        \begin{tablenotes} 
            \item[$\diamond$] Technically, since there is no PT on Librispeech, it is not appropriate to use the term “FT” as the model is directly trained on PRLVS or DATA2 full training data. However, we keep “FT” here for consistency. 
        \end{tablenotes}
    \end{threeparttable}
    \vspace{-5mm}
\end{table}

\vspace{-3mm}
\subsection{Few-shot Generalization}
Table \ref{tab:few_shot} presents the performance of both the baseline models and our proposed PMF-CEC in few-shot generalization capability. Here, ``0-shot" refers to rare words unseen during model training, whereas ``5-shot" and ``100-shot" indicate rare words that appeared five and one hundred times in the training set sentences, respectively. We collected rare words with different shots of occurrence. If a rare word is found in the corrected transcripts, it signifies that the word has been accurately corrected. This enables us to calculate the B-WER and RW-Recall for these rare words. ``Original” represents the baseline ASR system output without any AEC model applied, so its B-WER and RW-Recall vary under different rare word frequency conditions. On the test sets of Librispeech and PRLVS, both ED-CEC and PMF-CEC models demonstrated effective adaptation and handling of unseen or infrequently encountered rare words, showcasing promising zero-shot and few-shot learning capabilities. Additionally, our PMF-CEC model consistently outperformed the ED-CEC model, further demonstrating enhanced few-shot generalization capability through the integration of phoneme information.

\vspace{-3mm}
\subsection{Domain Adaptation with Limited Data}
To demonstrate the effectiveness of PMF-CEC in domains with limited training resources, we extracted 10\% of the data from the DATA2 and PRLVS training sets to finetune models trained on Librispeech 960-hour data. The specific results are detailed in Table \ref{tab:domain}. We evaluated three training strategies: 
\begin{itemize}[leftmargin=*]
\setlength{\itemsep}{0pt}
\setlength{\parsep}{0pt}
\setlength{\parskip}{0pt}
\item \textbf{FT (Full Data):} training from scratch on the complete dataset.
\item \textbf{PT + FT (10\% Data):} pretraining on Librispeech followed by finetuning with 10\% of the data.
\item \textbf{PT + FT (Full Data):} pretraining on Librispeech followed by finetuning with the entire dataset.
\end{itemize}

On one hand, the ``PT + FT (10\% Data)" strategy showed slightly higher WER and B-WER on PRLVS than the ``FT (Full Data)" strategy, whereas the opposite trend was observed on DATA2. However, both strategies significantly outperformed the original ASR transcripts on both datasets, indicating that finetuning pretrained models on limited amount of data yields substantial performance improvements in specific domains.
On the other hand, the ``PT + FT (Full Data)" strategy achieved further reductions in WER and B-WER, demonstrating optimal performance. This highlights that pretraining on ample data followed by finetuning with sufficient domain-specific data notably enhances model adaptation in specific domains. 
Additionally, the “PT” represents the setting where we directly evaluate the test sets of DATA2 and PRLVS using the pretrained model trained on the 960-hour LibriSpeech. As shown, this leads to a slight increase in WER compared with the original ASR transcripts, suggesting that our approach benefits from a small amount of domain-specific data for finetuning to better capture and adapt to local error patterns.

In conclusion, these experimental results validated the effectiveness of PMF-CEC in domains with limited training resources. Even with smaller datasets, the combined approach of pretraining and finetuning significantly improved performance, providing an effective pathway for domain adaptation.


\vspace{-3mm}
\subsection{Examples of Correcting Rare Words}
Fig. \ref{fig:example_visual} illustrates two specific examples of correcting rare words ``maier" and ``erlangen". In the first example, when the misrecognized word ``mayer" closely resembles the ground truth word ``maier" in spelling form, both CEC methods correctly select the appropriate word from the rare word list for correction. However, in the second example, where ``alum" is misrecognized as ``erlangen", despite their dissimilar spelling forms but similar pronunciation, the ED-CEC method's gate heatmap favors the output of the contextual decoder but incorrectly chooses the rare word ``algebra" similar to the spelling form of ``alum" for correction, resulting in a failed correction. 
In contrast, as illustrated by the heatmap, the PMF-CEC method accurately selected ``erlangen” from the rare word list to correct ASR errors by incorporating phoneme information, highlighting the necessity and effectiveness of leveraging pronunciation details to correct homophone errors.

Given that the rare words in Fig. \ref{fig:example_visual} are derived from lecture slides and lack significant phonetic ambiguity, Fig. \ref{fig:example_llm} presents a more challenging example specifically designed to evaluate the model’s ability to resolve homophone-induced errors. This case involves the rare named entity ``tsavo”, which is frequently misrecognized as the identically pronounced and more frequent word ``savo”.
In Fig. \ref{fig:example_llm}(a), both Whispering LLaMA and ED-CEC fail to correct the error. In contrast, PMF-CEC leverages phoneme-level information and contextual cues to accurately identify the correct rare word from the biasing list. Fig. \ref{fig:example_llm}(b) further compares MaLa-ASR, which fails to correct the error even though “tsavo” is explicitly included in the prompt, highlighting the limitations of text-only prompting strategies in resolving phonetic ambiguity. In contrast, PMF-CEC, as a lightweight postprocessing model, successfully corrects the error on the 1-best output from SLAM-ASR, achieving a strong balance between accuracy and efficiency.

\vspace{-3mm}
\subsection{Discussion}

\textbf{The Impact of Text Normalization.}
\label{sec:discussion}
We first investigate the effect of text normalization as a postprocessing step on the reference and hypothesis texts in the Whisper model.
Table \ref{tab:text_norm_result} presents the WER and B-WER results on the LibriSpeech test-clean and test-other datasets with text normalization applied.
As shown in the table, text normalization consistently lowers both WER and B-WER across all experimental settings. For example, in the test-clean dataset with the Whisper model, WER decreases from 4.00\% to 3.02\%, while B-WER drops from 15.08\% to 9.48\%. When incorporating the PMF-CEC model, WER is further reduced to 2.85\% and B-WER to 8.58\%.
Nevertheless, text normalization can mask actual improvements in biasing words. By standardizing formatting differences such as possessive contractions, numerical expressions, and British versus American spellings (e.g., fiber vs. fibre), text normalization may inadvertently correct errors in biasing words, thereby hindering a fair evaluation of contextual biasing methods. To achieve a more accurate assessment of contextual biasing effectiveness, evaluations should also be performed without applying text normalization.

\textbf{Phoneme Integration Beyond Postprocessing: Opportunities in LLM Decoding.} Our proposed PMF-CEC model incorporates phoneme information into the context-aware error correction module by jointly encoding textual and phonetic representations through a Context Encoder and a Context Phoneme Encoder. These multimodal representations are fused to improve the model’s ability to detect and correct homophone errors in ASR transcripts. However, this design remains within a postprocessing paradigm: phoneme information is only introduced after the ASR output has been generated and is used primarily to edit or refine the resulting text.

Since phoneme embeddings are not directly involved in the decoding process of the LM, their influence on sequence generation is delayed—that is, they do not participate in the real-time generation decisions. Although this modular structure is engineering-friendly and easy to integrate, it limits the extent to which phonetic information can interact with semantic modeling during text generation.

To further unlock the potential of phoneme-level features, future work may explore integrating phoneme embeddings earlier—directly into the decoding process of LLMs. This would enable the model to dynamically access and utilize phonetic cues at each generation step, leading to more precise contextual reasoning and improved disambiguation of phonetically similar words. Several mechanisms can be considered for this purpose, such as embedding-level fusion, cross-modal attention, or phoneme-guided prompting. These strategies will be investigated and evaluated in our future work.

\textbf{Why Seq2Seq (S2S) AEC Still Matters in the Era of LLMs.} Compared with LLM-based AEC and contextual biasing methods, S2S AEC models such as PMF-CEC offer a lightweight and efficient alternative for contextual AEC tasks. These models integrate seamlessly with standard 1-best ASR outputs, eliminating the need for prompt engineering and avoiding the high computational cost of generative inference. Moreover, S2S models (e.g., ED-CEC and PMF-CEC) remain robust when handling large biasing lists, whereas prompt-based LLM methods (such as MaLa-ASR) tend to degrade significantly as the size of the biasing list increases, limiting their applicability in practice. Notably, we also observe that combining the S2S-based PMF-CEC with generative AEC methods further improves correction performance, leveraging the fine-grained phoneme modeling of the former and the expressive semantic rewriting capabilities of the latter to achieve complementary benefits.
\begin{table}[t]
    \centering
    \caption{WER and B-WER on LibriSpeech test-clean and test-other sets using Whisper medium.en and PMF-CEC models, combined with \textbf{text normalization} (\%).}
    \vspace{-3mm}
    \label{tab:text_norm_result}
    \begin{tabular}{l|cccc}
        \toprule
        \multirow{2}{*}{\textbf{Method}} & \multicolumn{2}{c}{\textbf{Test-clean}} & \multicolumn{2}{c}{\textbf{Test-other}}  \\
        \cmidrule(lr){2-3} \cmidrule(lr){4-5} 
         & WER $\downarrow$ & B-WER $\downarrow$ & WER $\downarrow$ & B-WER $\downarrow$  \\
        \midrule
        \midrule
        Whisper & 4.00 & 15.08 & 6.83 & 21.89  \\
        \rowcolor[HTML]{EFEFEF} 
        \hspace{1em} + Text norm. & 3.02 & 9.48 & 5.84 & 19.88  \\
         Whisper + PMF-CEC & 2.99 & 9.61 & 6.02 & 16.45  \\
         \rowcolor[HTML]{EFEFEF} 
         \hspace{1em} \textbf{+ Text norm.} & \textbf{2.85} & \textbf{8.58} & \textbf{5.78} & \textbf{15.96}  \\
        \bottomrule
    \end{tabular}
    \vspace{-5mm}
\end{table}

\vspace{-2mm}
\section{Conclusion}
\label{sec:conclusion}
We improved upon our previous work to develop a phoneme-augmented context-aware error correction postprocessing method for E2E ASR models, referred to as PMF-CEC. By integrating phoneme information through multimodal fusion, our proposed PMF-CEC more effectively corrected homophone errors in ASR transcriptions, where the words sounded alike but had different spellings. Furthermore, we introduced a probability retention mechanism to boost the accuracy of the ASR error detection module. Experimental results demonstrated that our PMF-CEC outperforms our previous model and other contextual biasing methods. Moreover, despite the growing adoption of LLM-based ASR models, our lightweight and efficient postprocessing PMF-CEC remains valuable. PMF-CEC can be seamlessly integrated with existing ASR systems including LLM-based architectures without requiring model retraining or architectural modifications, making it especially suitable for latency-sensitive applications where real-time inference is critical.

\vspace{-2mm}




\bibliographystyle{IEEEtran}
\bibliography{IEEEtran}


 





\vfill

\end{document}